\documentclass[superscriptaddress,twocolumn,10pt,pra,aps,showpacs,longbibliography]{revtex4-2}

\usepackage{amsbsy}
\usepackage{amstext}
\usepackage{braket}
\usepackage[unicode=true,pdfusetitle,
 bookmarks=false,
 breaklinks=false,pdfborder={0 0 1},backref=false,colorlinks=false]
 {hyperref}
\hypersetup{
 bookmarksnumbered=false,bookmarksopen=false}

\usepackage{hyperref}
\hypersetup{colorlinks,citecolor=NavyBlue,filecolor=black,linkcolor=black,urlcolor=black}
\usepackage[dvipsnames]{xcolor}

\makeatletter

\@ifundefined{textcolor}{}
{%
 \definecolor{BLACK}{gray}{0}
 \definecolor{WHITE}{gray}{1}
 \definecolor{RED}{rgb}{1,0,0}
 \definecolor{GREEN}{rgb}{0,1,0}
 \definecolor{BLUE}{rgb}{0,0,1}
 \definecolor{CYAN}{cmyk}{1,0,0,0}
 \definecolor{MAGENTA}{cmyk}{0,1,0,0}
 \definecolor{YELLOW}{cmyk}{0,0,1,0}
}

\def\<#1>{\mathinner{\langle#1\rangle}}
\def\|#1>{\mathinner{|#1\rangle}}

\usepackage{amsmath}
\usepackage{graphicx}
\usepackage{amssymb}
\usepackage{txfonts,color}
\makeatother

\begin{document}

\title{Universal deterministic quantum operations in microwave quantum links}
\author{Guillermo F. Pe{\~n}as}\email{guillermof.pens.fdez@iff.csic.es}
\affiliation{Instituto de F{\'i}sica Fundamental, IFF-CSIC, Calle Serrano 113b, 28006 Madrid, Spain}
\author{Ricardo Puebla}
\affiliation{Instituto de F{\'i}sica Fundamental, IFF-CSIC, Calle Serrano 113b, 28006 Madrid, Spain}
\affiliation{Centre for Theoretical Atomic, Molecular and Optical Physics, Queen's University Belfast, Belfast BT7 1NN, United Kingdom}
\author{Tom\'as Ramos}
\affiliation{Instituto de F{\'i}sica Fundamental, IFF-CSIC, Calle Serrano 113b, 28006 Madrid, Spain}
\author{Peter Rabl}
\affiliation{Vienna Center for Quantum Science and Technology, Atominstitut, TU Wien, 1040 Vienna, Austria}
\author{Juan Jos\'e Garc\'ia-Ripoll}\email{juanjose.garcia.ripoll@csic.es}
\affiliation{Instituto de F{\'i}sica Fundamental, IFF-CSIC, Calle Serrano 113b, 28006 Madrid, Spain}
\begin{abstract}
    We propose a realistic setup, inspired by already existing experiments, within which we develop a general formalism for the implementation of  distributed quantum gates. Mediated by a quantum link that establishes a bidirectional quantum channel between distant nodes, our proposal works both for inter- and intra node communication and handles scenarios ranging from the few to the many modes limit of the quantum link. We are able to design fast and reliable state transfer protocols in every regime of operation, which, together with a detailed description of the scattering process, allows us to engineer two sets of deterministic universal distributed quantum gates. Gates whose implementation in quantum networks does not need entanglement distribution nor measurements. By employing a realistic description of the physical setup we identify the most relevant imperfections in the quantum links as well as optimal points of operation with resulting infidelities of $1-F \approx 10^{-2}-10^{-3}$.
\end{abstract}
\maketitle

\section{Introduction}


Near term quantum processors are constrained in size because of hardware limitations: density of packing, trapping and control lines;  cross-talk  between qubits; cooling power; etc. A near term solution to this problem is the creation of quantum links between quantum processors\ \cite{Kurpiers2017, Leung2019, Chang2020}. Without the ambition and complexity of a quantum internet\ \cite{Kimble2008,Wehner2018}, such coherent channels enable larger scale distributed quantum computations (DQC)~\cite{Cirac1999} and the interconnection of quantum processors with auxiliary components such as sensors and memories. Quantum state transfer, introduced by Cirac {\em et al.}~\cite{Cirac1996} is a central idea in the design of coherent quantum links. It enables the deterministic exchange of quantum information by the controlled emission and perfect absorption of photons at the extremes of a photonic link, as demonstrated in the optical~\cite{Ritter2012} and microwave regime~\cite{Kurpiers2017}, in a way that is arguably robust against noise in the link~\cite{Cirac1999,Xiang2017,Vermersch2016}. Building on this and the notion of a photonic link, there have been proposals and realizations that support entanglement distribution and two-qubit gates using adiabatic protocols~\cite{Pellizzari1997,Chen2007,Ye2008,Clader2014,Vogell2017,Leung2019}, photon mediated couplings~\cite{Serafini2005,Yin2007,Lu2008,Hua2015, Bienfait2019}, chiral propagation~\cite{Ramos2016,Calajo2019}, quantum teleportation~\cite{Fedorov2021} and non-deterministic operations~\cite{Bose1999, Dickel2018, Narla2016, Pfaff2014, Nolleke2013, Bao2012, Olmschenk2009}  (see~\cite{Northup2014} and references therein).

In this work we address the implementation of photon-based state transfer~\cite{Cirac1996} and photon-cavity scattering~\cite{Duan2004} as primitives for a distributed deterministic quantum toolbox that works in commercial non-chiral microwave quantum links (QL)~\cite{Kurpiers2017,Leung2019,Chang2020}. We demonstrate numerically protocols that implement both primitives for real waveguides with diffraction and non-uniform couplings, under very different limits of length and photon bandwidth---from a continuum limit, to a situation where only one or few modes participate. In all cases, we show that the dynamics is very well approximated by a continuum limit input-output theory, which not only provides fast controls---an order of magnitude faster than adiabatic protocols even for short links~\cite{Vogell2017, Chang2020}---but also quantifies the errors due to non-linear dispersion relations, few-mode dynamics and point-scattering diffraction. Our study provides a unified theoretical framework that can be used to engineer quantum links inside the same refrigerator~\cite{Chang2020} as well as long distance links~\cite{Kurpiers2017,Magnard2020}.

Using this framework we can also design two different implementations of a universal two-qubit quantum gate working on quantum processors connected by a quantum link. The first implementation is based on quantum state transfer~\cite{Cirac1999}, while the second one combines this primitive with photon-cavity scattering gates~\cite{Duan2004,Kono2018}. Both gates are fast and bandwidth-limited and exhibit very low near-term infidelities $1-F\sim 10^{-2}-10^{-3}$. The gates may thus be used to implement quantum algorithms distributedly among quantum processors. In this respect, they represent a more viable near-term alternative than DQC algorithms based on entanglement distribution and measurements, whose quality will be severely limited by the poor measurement fidelities in superconducting circuits. Our results highlight the suitability of pulse shaping protocols derived from a continuum limit to perform universal two-qubit quantum gates across a wide range of distances and parameters, and provide insight on the main limitations posed by realistic dispersion relations. This protocol is not typically considered in this few mode regime and, as we will show, it outperforms all other methods with which we compared it.

Finally, it must be remarked that the results in this work are not constrained to superconducting circuits, but generalize straightforwardly to other quantum optical platforms with non-ideal quantum links. This includes phononic and hybrid quantum computer designs~\cite{Stannigel2010, Stannigel2011, Wang2011, Tian2012, Lemonde2018,Calajo2019} and other microwave photon links~\cite{Brecht2016, Fedorov2021}, where our tools can be used to implement quantum network operations in both short length quantum buses, as well as longer communication lines.

This work is organized as follows. In Sec.~\ref{sec:setup} we introduce a general quantum optical Wigner-Weisskopf model for two quantum processors connected by a generic bosonic quantum link. The model considers a realistic dispersion relation and frequency dependent couplings between the qubits and the link. In Sec.~\ref{sec:QST} we use this model to study the photon-based state transfer~\cite{Cirac1996} between two processors in this setup. We show how a single control can operate both in the long-distance and short-distance limits, explaining the imperfections due to diffraction and few-mode dynamics. In Sec.~\ref{sec:scatt} we discuss the entangling gate between a travelling photon and a qubit-hosting cavity~\cite{Duan2004} under conditions of limited photon bandwidth. Finally, in Sec.~\ref{sec:applications} we put this theory to work in particular applications. We first explain how our results influence the design of short and long quantum links. Then, in Sec.~\ref{ss:QGT} we propose a protocol for implementing a quantum gate based purely on state transfer. Finally, Sec.~\ref{ss:cphase} introduces a design for a \textit{passive} two-qubit quantum gate based on photon emission and reabsorption, and scattering with a second quantum node. Section~\ref{sec:summary} summarizes our results and provides an outlook to open questions.

\section{Setup}\label{sec:setup}

\begin{figure}
    \includegraphics[width=\columnwidth]{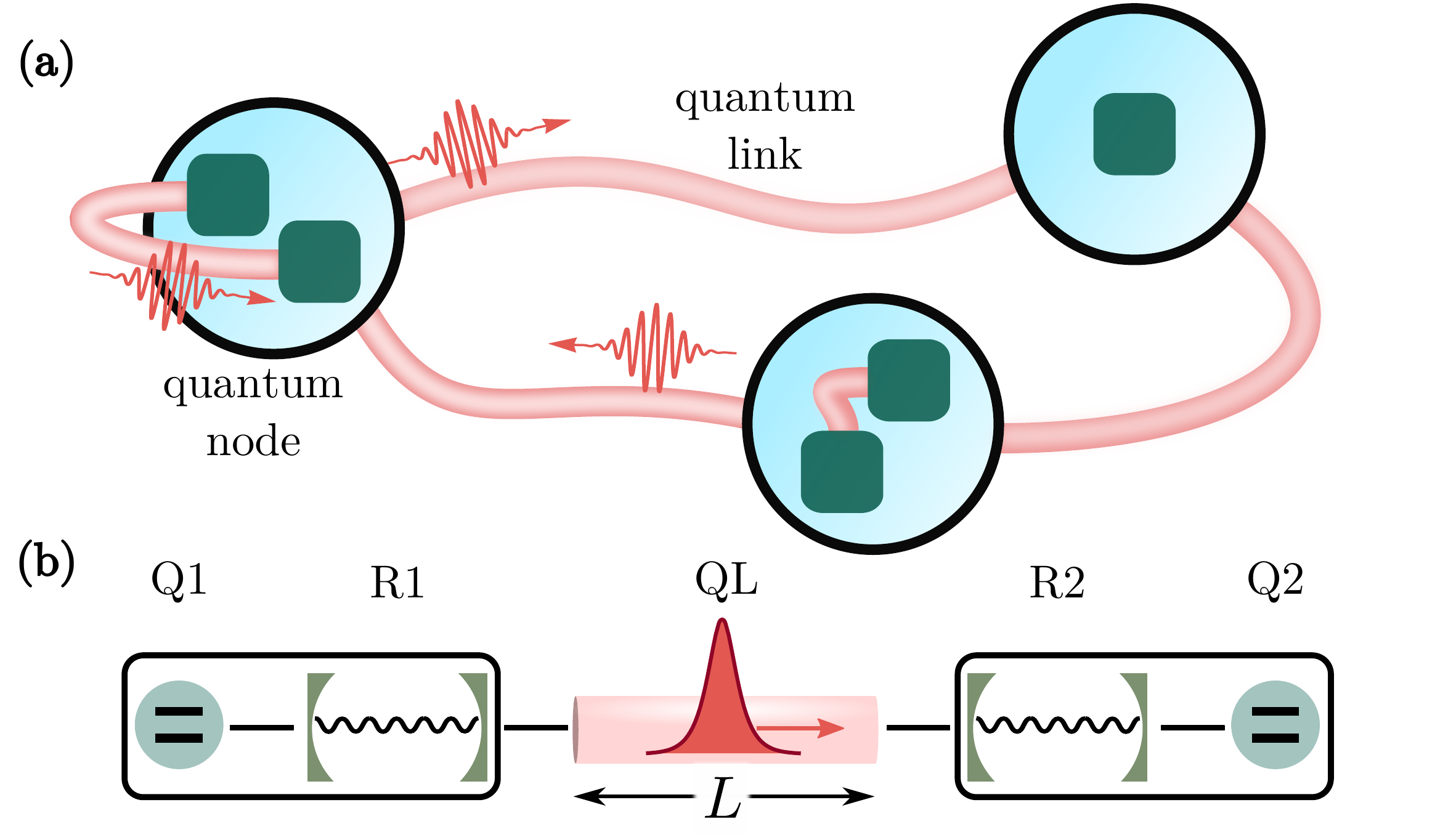}
    \caption{(a) Sketch of a quantum network, where multiple quantum nodes connected via quantum links (QL) of different lengths allowing for a inter- and intra-node bi-directional exchange of quantum information via photon pulses. (b) Building block of the physical setup, where two quantum nodes, consisting of a qubit coupled to a resonator, Q1 and R1 and Q2 and R2, respectively, are connected via a waveguide of length $L$ that acts as a QL (cf. Eq.~\eqref{eq:H}).}
    \label{fig1}
\end{figure}

The contributions in this work aim towards a vision sketched in Fig.~\ref{fig1}(a), in which photonic waveguides of various sizes interlink both quantum computing nodes as well as quantum devices withing a single node. The building block in these quantum links (cf. Fig.~\ref{fig1}(b) and Refs.~\cite{Kurpiers2017,Magnard2020}) will be a waveguide that connects cavities in different nodes, each one hosting a single qubit. The waveguide has a length $L$. It is not chiral and acts as a bidirectional link between both qubits, which is where the information will be ultimately stored and used. We model the link with the Hamiltonian ($\hbar=1$)
\begin{align}\label{eq:H}
  H&=H_{\rm QL}+\sum_{j=1,2}\left[ H_{\rm N_j}+ H_{\rm N_j-QL}\right],\mbox{ with}\\
  H_{\rm N_j}&=\delta_j \sigma^+_j\sigma_j^-+\Omega_{{\rm R}j} a^\dagger_j a_j+g_j(t)\left(\sigma^+_ja_j +{\rm H.c.}\right),\notag\\
   H_{\rm QL}&=\sum_m \omega_m b_m^\dagger b_m,\notag\\
   H_{\rm N_j-QL}&=\sum_m G_{m,j} \left(b_m^\dagger a_j+{\rm H.c.}\right).\notag
\end{align}
Each quantum node contains a qubit and a cavity $H_{\text{N}_j}$ with a coupling $g_j(t)$ that can be tuned for wavepacket engineering~\cite{Kurpiers2017,Magnard2020}. The waveguide $H_\text{QL}$ hosts a family of modes with frequencies $\omega_m$, which connect both cavities through a static coupling $H_{\rm N_j-QL}$. The $a_j$ and $b_m$ are Fock operators, and we use spin-$1/2$ Pauli matrices $\sigma_j^+=\ket{1}_j\!\bra{0}_j$ for the qubits.

Our study assumes a standard WR90 rectangular microwave guide~\cite{Pozar}, as it is the case in the experiment reported in~\cite{Magnard2020}. The ${\rm TE_{10m}}$ modes frequencies $\omega_m=c\sqrt{(\pi/l_1)^2+(m\pi/L)^2}$, depend on the vacuum speed of light $c$,  the broad wall dimension $l_1=2.286$ cm and the wavenumber $k_m=m\pi/L$. The resonator-waveguide coupling amplitude can be deduced from each cavity's decay rate $\kappa_j$ and the parity of the stationary modes as $G_{m,j}=(-1)^{m(j-1)}\sqrt{\kappa_j v_g \omega_m/(2\Omega_{\text{R}j} L)}$. We estimate the group velocity $v_g=\mathrm{d}\omega(k)/\mathrm{d}k$ around the frequency closest to both cavities $\omega_c=\omega(k_c)\simeq\Omega_{\text{R}1}=\Omega_{\text{R}2}$, which we assume identical. We choose a gauge in which $g_j, G_{m,j}\in \mathbb{R}$, and impose a hierarchy of interactions $|\Omega_{{\rm R}j}-\delta_j|\ll |\Omega_{{\rm R}j}+\delta_j|$, and $|g_j|,|\kappa_{j}|\ll \omega_m,\Omega_{{\rm R}j},\delta_j$ which justifies the rotating-wave approximation used in $H_{\rm N_j}$ and $H_{\rm N_j-QL}$.

We choose to work at the X-band of the waveguide ($8-12$ GHz) as it is usual in experiments \cite{Magnard2020,Kurpiers2017}. For typical parameters $\delta_{1,2},\omega_{1,2}\sim 2\pi\times 8.4$ GHz, $\kappa_j\sim
\text{MHz}$, the group velocity of the central mode is $v_g = c \sqrt{1- \left(\pi c/a\omega \right)^2} \approx 2c/3$. A relevant parameter is the free spectral range (FSR) $\Delta\omega_m=\omega_{m+1}-\omega_{m}\sim \text{MHz}$, which depends on $L$. The interplay between $\kappa_j$ and $\Delta\omega$ will prove relevant in later developments.

Since we are interested in operations at cryogenic temperatures, thermal excitations can be neglected. Also, the protocols we describe use at most one travelling photon, thus, we study the link using the single-excitation Wigner-Weisskopf ansatz,
\begin{equation}
  \ket{\Psi(t)} = \left[\sum_{j=1,2}\left(q_j(t)\sigma_j^++c_j(t)a_j^\dagger\right)+\sum_m \psi_m(t)b_m^\dagger \right]\ket{\mathbf{0}}.
\end{equation}
This normalized state describes an excitation created on top of the trivial ground state $\ket{\mathbf{0}}$ of Eq.~\eqref{eq:H}. The ansatz evolves according to the coupled ordinary differential equations
\begin{align}\label{eq:eom1}
    i\dot{q}_j(t)&=\delta_j q_j(t)+g_j(t) c_j(t),\\ \label{eq:eom2}
    i\dot{c}_j(t)&=\Omega_{{\rm R}j} c_j(t)+\sum_m G_{m,j}\psi_m(t)+g_j(t) q_j(t),\\ \label{eq:eom3}
    i\dot{\psi}_m(t)&=\omega_m \psi_m(t)+\sum_{j=1,2}G_{m,j}c_j(t).
\end{align}
This description is valid for all regimes of operation and all times. Yet, when photons are shorter than the waveguide, one may resort to input-output relations that describe the emission of the photons by the qubit, or the scattering of photons by the cavities. These are the models supporting deterministic state transfer~\cite{Cirac1996} and collisional gates~\cite{Duan2004}.

\section{Quantum state transfer}\label{sec:QST}


\subsection{Ideal protocol}

The deterministic quantum state transfer refers to the physical exchange of quantum information between remote qubits according to the following operation
\begin{equation}\label{eq:ST}
(\alpha \ket{0}_1+\beta\ket{1}_1)\otimes\ket{0}_2 \xrightarrow{\mathcal{T}} \ket{0}_1\otimes (\alpha \ket{0}_2+\beta\ket{1}_2),
\end{equation}
originally proposed in Ref.~\cite{Cirac1996}. As discussed in Ref.~\cite{Cirac1999} and demonstrated experimentally in~\cite{Ritter2012}, the state transfer operation can be mediated by a single photon. In this protocol, the possible excitation $\ket{1}_1$ of qubit Q1 is transferred to a propagating photon that is perfectly absorbed by the cavity R2 and the qubit Q2. To engineer this perfect absorption, the couplings $g_1(t)$ and $g_2(t)$ are modulated in time, creating a time-reversal symmetric wavepacket that is emitted by Q1-R1 and absorbed by R2-Q2 with near to $100\%$ probability.

In the literature, the derivation of this protocol relies on a formalism that does not take into account the length of the waveguide, because the photon is much shorter. Alternatively, this is a limit in which the bandwidth of the photon is broader than the free-spectral range $\Delta\omega_m$, and we can apply a continuous limit, input-output theory to describe the generation and absorption of the photon. However, in the following sections we will provide evidence that perfect state transfer works also in the limit of very short quantum links, even when few discrete modes participate in the dynamics, without any relevant changes in the controls or the dynamics.

When we derive the state transfer in the usual way~\cite{Cirac1996}, the coupling $g_1(t)$ determines the shape of the emitted photon. The coupling $g_2(t)$ is designed to cancel the probability that the photon is reflected, which makes this control the time-reversed variant of $g_1(t)$. In this work we follow the standard controls that shape the photon with a sech profile $\ket{\xi}=\int d\omega f(\omega)b^\dagger_\omega\ket{{\bf 0}}$ with $f(\omega)=\sqrt{\pi/(2\tilde{\kappa})}{\rm sech}(\pi(\omega-\omega_c)/\tilde{\kappa})$. These photons maximize the frequency bandwidth provided by the cavity $\tilde\kappa\leq \kappa_1$. They have a temporal width $\sigma_t=\pi/(\sqrt{3}\tilde{\kappa})$ and are generated by the qubit-resonator coupling $g_1(t)=g(t+t_d/2;\tilde\kappa,\kappa_1)$ (cf. App.~\ref{app:pulse})
\begin{align}\label{eq:gt}
    g(t;\tilde\kappa,\kappa)=\frac{\kappa-\tilde{\kappa}\tanh(\tilde{\kappa} t/2)}{2\sqrt{ (1+e^{-\tilde{\kappa} t})\kappa/\tilde{\kappa}-1}},
\end{align}
which we center around $t=-t_d/2$. Under ideal conditions, the photon is perfectly absorbed by the second resonator R2 when $g_2(t)=g(-t+t_d/2;\tilde\kappa,\kappa_2)$ and $\tilde\kappa\leq \kappa_2.$ Note that to implement the protocol as fast as possible, we have selected the maximum value of the qubit-resonator coupling $\text{max}\{g_{1,2}\}= \kappa_{1,2}/2$. This makes the long photon and few modes limits equivalent. The delay between controls is adjusted to match the propagation time of light between cavities $t_d=t_p\equiv L/v_g$, and the whole protocol runs over a time $t\in[-T,T]$ long enough to prevent significant truncations of the sech-pulse control---in our case, $g_1(-T),g_2(T)\lesssim 2\pi \times 10^{-5}$ MHz. To ensure this, we make the duration dependent on the intrinsic time-scale of the problem by means of $2T = t_p + 10/\tilde{\kappa}$. As we will show later, such small initial values for the couplings are indeed not required to reach high fidelities.

It is worth mentioning that the control~\eqref{eq:gt} may compensate asymmetries in the cavity properties $\kappa_2\neq \kappa_1$ in a straightforward manner. In addition, one may generate photons with a bandwidth reduced by a factor $\eta=\kappa_1/\tilde\kappa\geq 1$. As we will discuss later, this is relevant since narrower photons are less prone to imperfections. However, since such narrow photons demand an $\eta$ times longer protocol, there is a trade off between imperfections and decoherence. In addition, it is worth noting that a limited bandwidth with typical values (i.e. few hundreds of MHz) does not introduce a significant distortion of the control $g(t)$ (see App.~\ref{app:Bandwidth} for details). 

\subsection{Experimental limitations and imperfections}\label{ss:imp}

A careful study of this process in a superconducting architecture reveals two important effects. The first one is the appearance of Lamb shifts~\cite{DiazCamacho2015}, caused by (i) the non-uniform coupling between the cavity and the waveguide $G_{m,j}\propto \sqrt{\omega_m}$ and (ii) the curvature in the dispersion relation (cf. Sec.~\ref{sec:setup}). Because of this, the qubit interacts with a dressed resonator of frequency $\Omega_{\text{R}j}+\lambda_j$, which is the measurable resonator frequency. Theoretically, this can be compensated by detuning the qubit by a similar amount $\tilde{\delta}_j=\delta_j+\lambda_j$ to achieve a resonant condition between qubit and resonator. As an example,  we numerically find $\lambda_{1,2}=-0.0116\kappa$ for a waveguide of $L=30\text{m}$. We have also checked analytically, following \cite{DiazCamacho2015} and \cite{Gely2017}, that there are no relevant cutoff effects and that the Lamb shifts do not diverge in any of the situations considered throughout this article. Moreover, this does not constitute a problem in real experiments since the frequency that gets detected is the one with the shift included.

The second effect is the distortion of the wavepacket caused by the waveguide's non-linear dispersion relation, which reduces the probability of absorption by the second node. We quantify this distortion by the overlap $|z|^2=\left|\langle \tilde{\xi}(t_p)|\xi(t_p)\rangle\right|^2$ between the distorted photon $\ket{\xi(t_p)}$ that has propagated a time $t_p$ in the real waveguide, and the ideal photon $\ket{\tilde{\xi}(t_p)}$ that would have experienced a similar dynamic in a waveguide with  a linear dispersion relation. This leads to (cf. App.~\ref{app:dis})
\begin{align}\label{eq:stfidelity}
|z|^2&=\left|\int d\omega |f(\omega)|^2 e^{-it_p(\omega-\omega_c)^2 D /(2v_g^2)}\right|^2\approx 1-\frac{L^2D^2\kappa^4}{180v_g^6 \eta^4},
\end{align}
where we have introduced the curvature of the dispersion relation, $D=\mathrm{d}^2\omega(k)/\mathrm{d}k^2$ at the central frequency $\omega_c$ (in our case $D\approx 10^6\text{m}^2/\text{s}$ for the standard WR90 waveguide used in~\cite{Magnard2020}). This formula quantifies the distortion suffered by the photon and also the reduction in the probability of state transfer. Note how it scales with the photon bandwidth $1-|z|^2\propto \kappa^4/\eta^4$ (see App.~\ref{app:dis} for further details), indicating that narrow photons cannot ``see'' the curvature of the waveguide's dispersion relation.

In addition, the main source of decoherence in these systems stems from the qubit decay given by $T_1$. In order to quantify this effect, we compute the time during which the qubits are populated, namely, $\tau=\int dt (|q_{1}(t)|^2+|q_2(t)|^2)$, integrated over the whole protocol duration. In this manner, the fidelity is reduced by a factor $e^{-\tau/T_1}$.

\begin{figure}[t]
    \includegraphics[width=\columnwidth]{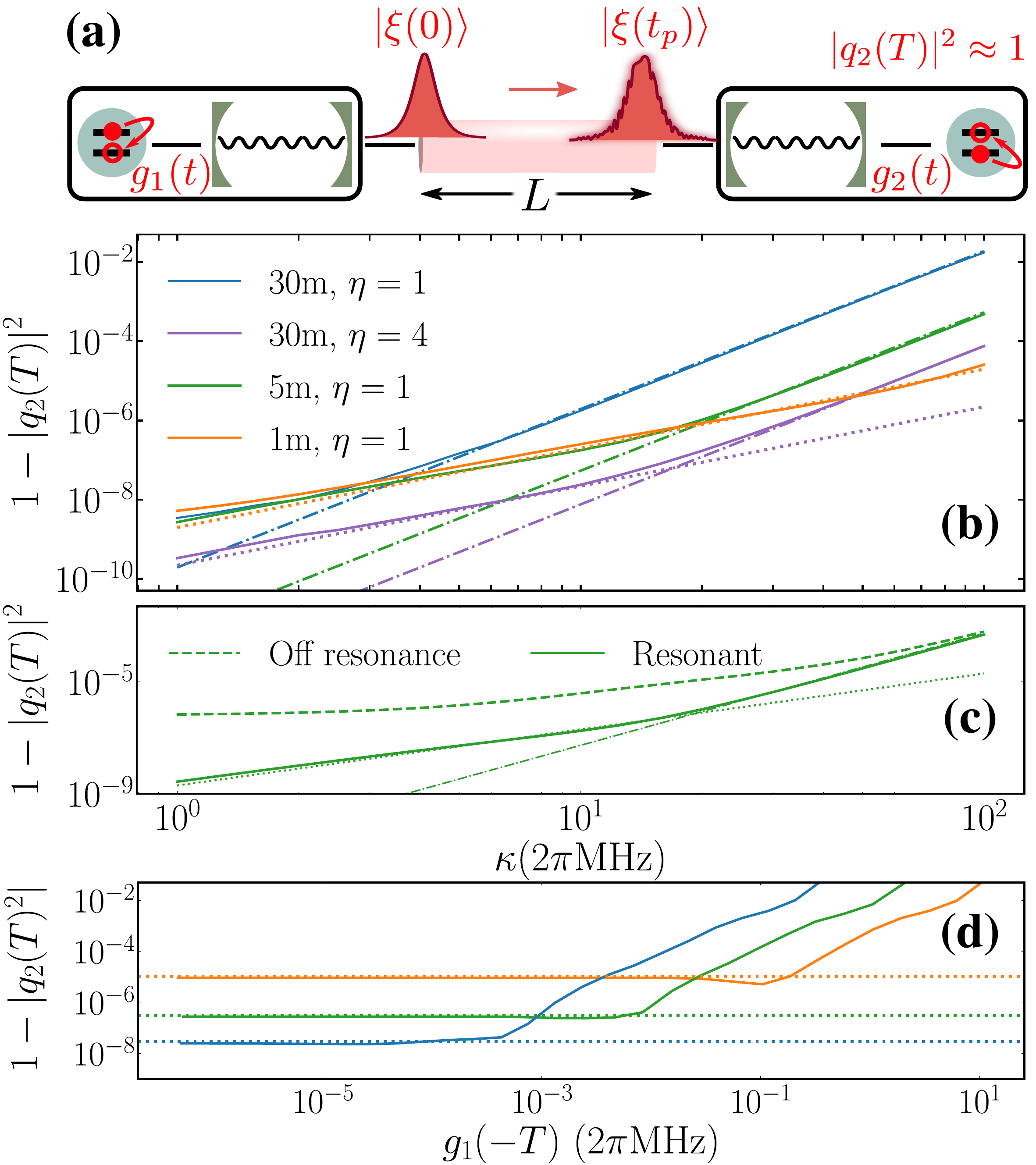}
        \caption{(a) Schematic illustration of the quantum state transfer protocol: By applying the controls $g_{1,2}(t)$ one can transfer the excitation from Q1 to Q2 via a photon propagating through the QL that distorts it, and thus, $|q_2(T)|^2\approx 1$ at the end of the protocol. Panel (b) shows the quantum state transfer efficiency $1-|q_2(T)|^2$ as a function of $\kappa$ for three different QL lengths ($L=30$, $5$ and $1$ m), and distinct photon bandwidths, $\eta=1$ and $\eta=4$. The dashed-dotted lines correspond to the limit imposed by diffraction, cf. Eq.~\eqref{eq:stfidelity} (short photon limit). The dotted lines represent the length-independent limit (long photon limit). The crossover between these regimes takes place when $2\sigma_t\approx t_p$, i.e., $\kappa/\eta\approx 2\pi v_g/(\sqrt{3}L)$.
    Panel (c) shows the impact in quantum state transfer efficiency when the resonant condition is not met, that is, $\Omega_{\rm R1}=\Omega_{\rm R2}=\omega_c$ (resonant) versus $\Omega_{\rm R1,R2}=\omega_c+\Delta\omega_c/2$ (off resonant), i.e., R1 and R2 lie in between two ${\rm TE}_{10m}$ modes. Panel (d) shows the impact that a truncation of the control $g(t)$ has on the quantum state transfer efficiency, for $\kappa=2\pi\times 90, 18, 3 $ MHz and $L=1$, $5$ and $30$ m, respectively (same style as in (b)). This indicates that the diffraction limit can be reached already with $g_1(-T)\approx 2\pi\times 10^{-3}$ or $10^{-1}$ MHz, depending on the parameters. }
    \label{fig2}
\end{figure}

\subsection{Realistic simulations}

We have simulated the complete state transfer protocol $\mathcal{T}$, using the model \eqref{eq:H} with a converged number of modes, starting from $q_1(-T)=1$ and characterizing the transfer efficiency of the protocol by the population of the second qubit $|q_2(T)|^2$ at the end of the numerical simulation. Fig.~\ref{fig2}(b) shows the efficiency as a function of the bandwidth of the cavity $\kappa=\kappa_{1,2}$ for different waveguide lengths, $L=1, 5$ and $30$ m, focusing on photons with the maximum bandwidth $\tilde{\kappa}=\kappa$, but also including narrowed photons $\tilde{\kappa}=\kappa/4$ ($\eta=4$) for the $30$ m case.

When the temporal spread of the photon $\sigma_t$ is smaller than the propagation time $t_p$, i.e. for $\kappa/\eta\gtrsim 2\pi v_g/(\sqrt{3}L)$, the photon fits completely within the QL. In this \textit{short photon limit} the transfer efficiency is limited by the diffraction along the waveguide $|q_2(T)|^2\approx |z|^2$ (cf. Eq.~\eqref{eq:stfidelity}). Since this value decreases rapidly with $\eta$, choosing narrower photons $\tilde{\kappa}<\kappa$ ($\eta>1$) reduces the infidelity by a factor $\eta^{-4}$, while slowing down the transfer and increasing the protocol duration $2T$ by a factor $\eta$ (cf. Fig.~\ref{fig2}(b)).

In the \textit{long photon limit} $\kappa/\eta\lesssim 2\pi v_g/(\sqrt{3}L)$, the wavepacket is never fully contained within the waveguide and the input-output theory that justifies the control~\eqref{eq:gt} is no longer valid. In this limit, the notion of photon propagation disappears and the dynamics is described by a few dressed photonic modes (cf. App.~\ref{app:hyb}). Indeed, this regime has been typically considered unsuited for pulse shaping protocols since the photon only populates one (or very few) of the waveguide modes.

Contrary to this common belief,  we have found excellent results for the quantum state transfer even in the long photon limit, without any change in the controls $g_{1,2}(t)$ (cf. Eq.~\eqref{eq:gt}). Using these controls, there is a limit to the state transfer efficiency which we attribute to Stark shifts that hinder a perfect emission and absorption of the excitation (cf. App.~\ref{app:hyb}). This model predicts an infidelity $1-|q_2(T)|^2\sim {\kappa}^2/\eta^2$ that does not depend on the waveguide length, as confirmed by the dotted line in Fig.~\ref{fig2}(b) for $\eta=1$ and $L=30\text{m}, 5\text{m}, 1\text{m}$. Finally, it is worth noting that, owing to the nature of this regime and due to the small values obtained for $1-|q_2(T)|^2$, the transfer efficiency is very sensitive to the resonant condition between R1, R2 and the closest mode of the waveguide $\Omega_{R1}=\Omega_{R2}=\omega_c$. Indeed, as shown in  Fig.~\ref{fig2}(c), when the cavities do not fulfill this relation, the protocol saturates at much higher infidelity.

In addition, we investigate the role  that the truncation of the sech-pulse plays in the resulting quantum state transfer efficiency. For that we decrease the protocol duration time $2T$ such that $g_1(-T)$ increases from $2\pi\times 10^{-5}$ MHz to few MHz. As shown in Fig.~\ref{fig2}(d), we find that below a certain value for $g_1(-T)$ the quantum state transfer efficiency saturates to the diffraction limit. This clearly indicates that the condition $g_1(-T)\lesssim 2\pi \times 10^{-5}$ MHz can be lifted, and that the experimentally reported values in the range of $10^{-2}$ or $10^{-1}$ MHz~\cite{Magnard2020} may be sufficient to reach high-fidelity quantum operations.

\subsection{Comparison with adiabatic and direct SWAP protocols}\label{ss:stirap_swap}

\begin{figure}
  \includegraphics[width=\columnwidth]{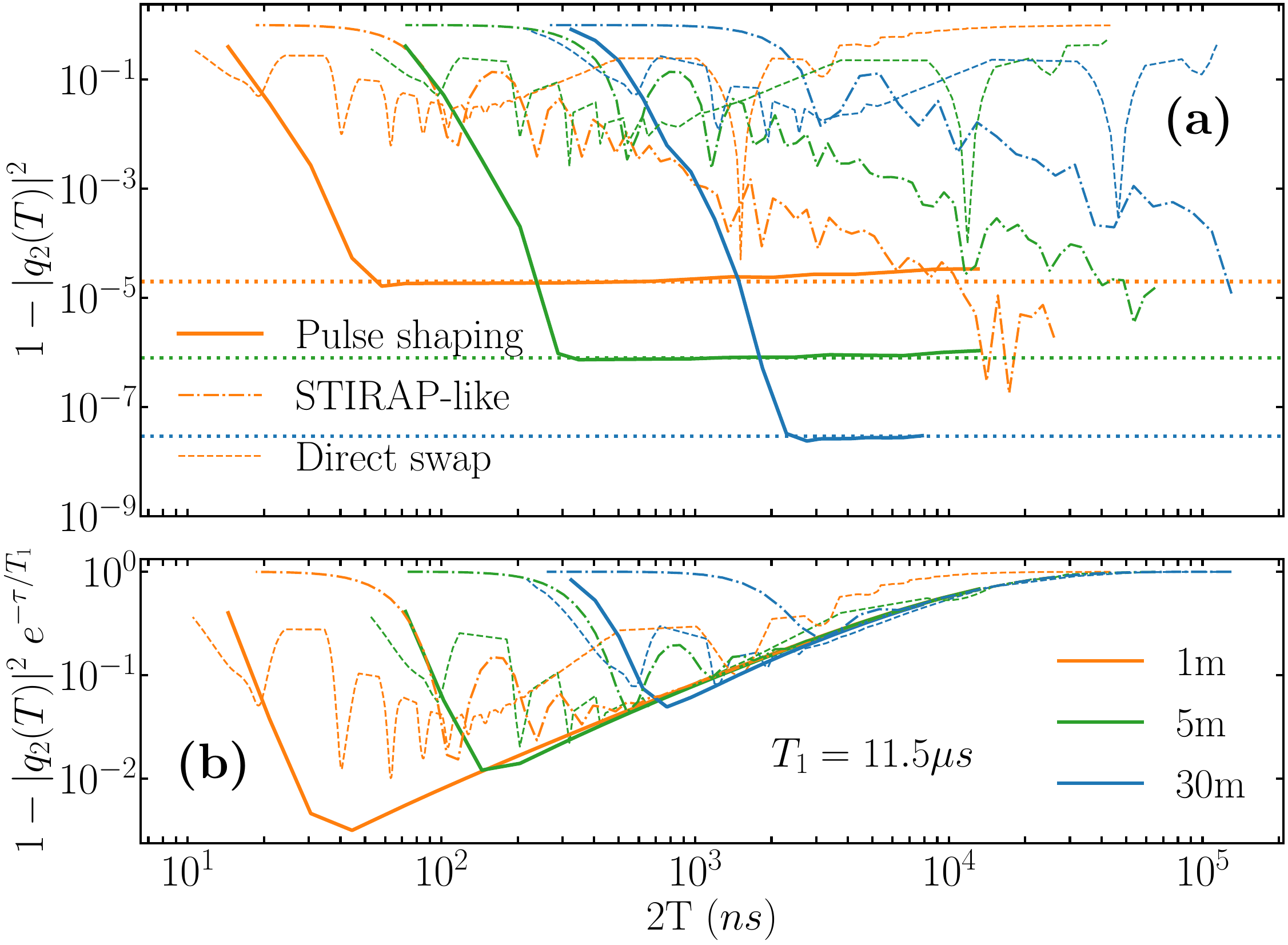}
  \caption{(a) Comparison of the quantum state transfer efficiency $1-|q_2(T)|^2$ as a function of the protocol duration for the direct swap, STIRAP-like and pulse shaping following Eq.~\eqref{eq:gt}, for $L=1$, $5$, and $30$ m; and $\Delta\omega_c \approx \kappa = 2 \pi \times 90$, $18$ and $3$ MHz, respectively. The horizontal dotted lines correspond to the efficiency limit posed by the wavepacket distortion (cf. Eq.~\eqref{eq:stfidelity}), to which the pulse shaping saturates. (b) State transfer efficiency for the same three protocols but including a finite coherence time of the qubits of $T_1= 11.5\mu s$, following Ref.~\cite{Chang2020}. The decoherence has been introduced according to $1-|q_2(T)|^2e^{-\tau/T_1}$.}
  \label{fig3}
\end{figure}

The excellent performance of the pulse shaping method in the long photon limit can be compared with a state-of-the-art adiabatic STIRAP-like protocol used in previous works~\cite{Vogell2017,Chang2020, Leung2019} and with a direct-SWAP gate, as reported in~\cite{Serafini2005}. The STIRAP-like  technique mimics the methods known from atomic physics to transfer excitations between $\Lambda$ configuration atoms~\cite{Vitanov2017}.  Such adiabatic protocols may be implemented by $g_1(t)=g_0\sin((t+T)\pi/(4T))$ and $g_2(t)=g_0\cos((t+T)\pi/(4T))$, with $g_1(-T)=0$ and $g_2(-T)=g_0$, as done in~\cite{Chang2020}. This achieves perfect state transfer from Q1 to Q2 in the adiabatic limit $T\gg 1/g_0$. To the contrary, a direct-SWAP gate consists in an  always-on and constant couplings $g_{1,2}=g$ that approximately realizes a quantum state transfer in a time $g t\approx \pi$~\cite{Serafini2005}.

Following~\cite{Chang2020} we choose the resonators decay $\kappa = \kappa_{1,2} = \Delta\omega_c$ and $g_0=\kappa/5.26$  (this value of $g_0$ is imposed by the adiabatic condition, see supplemental material of \cite{Chang2020}), and compare our results obtained under~\eqref{eq:gt} with those using a STIRAP-like protocol as the time of the protocol increases (cf. Fig.~\ref{fig3}(a)). As expected, the pulse shaping efficiency saturates to a certain value due to the distortion of the wavepacket (cf. Eq.~\eqref{eq:stfidelity}), yet this value is reached for protocol speeds that are two orders of magnitude shorter than STIRAP-like ones. For comparison, we also include in Fig.~\ref{fig3}(a) the results of a direct-SWAP protocol, where  we scan different values of $g$ and find the optimal protocol time that maximizes the quantum state transfer. Yet, while pulse shaping and STIRAP protocols are robust against time variations, the efficiency of a direct-SWAP is very sensitive to deviations from the optimal time and still do not offer an improvement over pulse shaping. See App.~\ref{app:swap} for more details regarding the direct-SWAP and an improved STIRAP-like protocol.

This difference is very relevant when we consider imperfections. Indeed, even though STIRAP is not affected by diffraction and approaches perfect transfer for long enough controls, the fact that the protocol is much slower means that it will be more sensitive to decoherence. As example, Fig.~\ref{fig3}(b) considers a realistic protocol with all imperfections and a finite qubit lifetime $T_1=11.5\mu$s, comparable to the one in state-of-the-art experiments~\cite{Chang2020}. When decoherence is taken into account, the order of magnitude difference in the protocol duration translates into more than an order of magnitude improvement in the protocol fidelity. In addition, although a direct-SWAP protocol can reach higher fidelities than STIRAP in faster times, it still does not improve the resulting fidelity of pulse shaping and it becomes sensitive to deviations from an optimal operation time, cf. App.~\ref{app:swap}.

\section{Photon phase via scattering}\label{sec:scatt}

In addition to state transfer, the second important primitive that has been introduced for creating photonic networks is the photon-qubit entangling operation by Duan and Kimble~\cite{Duan2004}. In this protocol, a propagating photon is reflected by a cavity whose frequency depends on the state of an off-resonantly coupled qubit. The reflected photon carries a phase that can be used to implement different protocols, as explained in Sec.~\ref{ss:cphase}.

The setup for the scattering phase is shown in Fig.~\ref{fig4}(a). A single photon is created by the qubit Q1 in the first node. This photon propagates and interacts with the second cavity R2, where it enters and exits without distortion, only acquiring a phase shift. We will first analyze this phase in the limit of very long wavepackets or quasimonochromatic photons, as a function of the photon-cavity detuning.

\subsection{Quasi-monochromatic scattering}

\begin{figure}
    \includegraphics[width=\columnwidth]{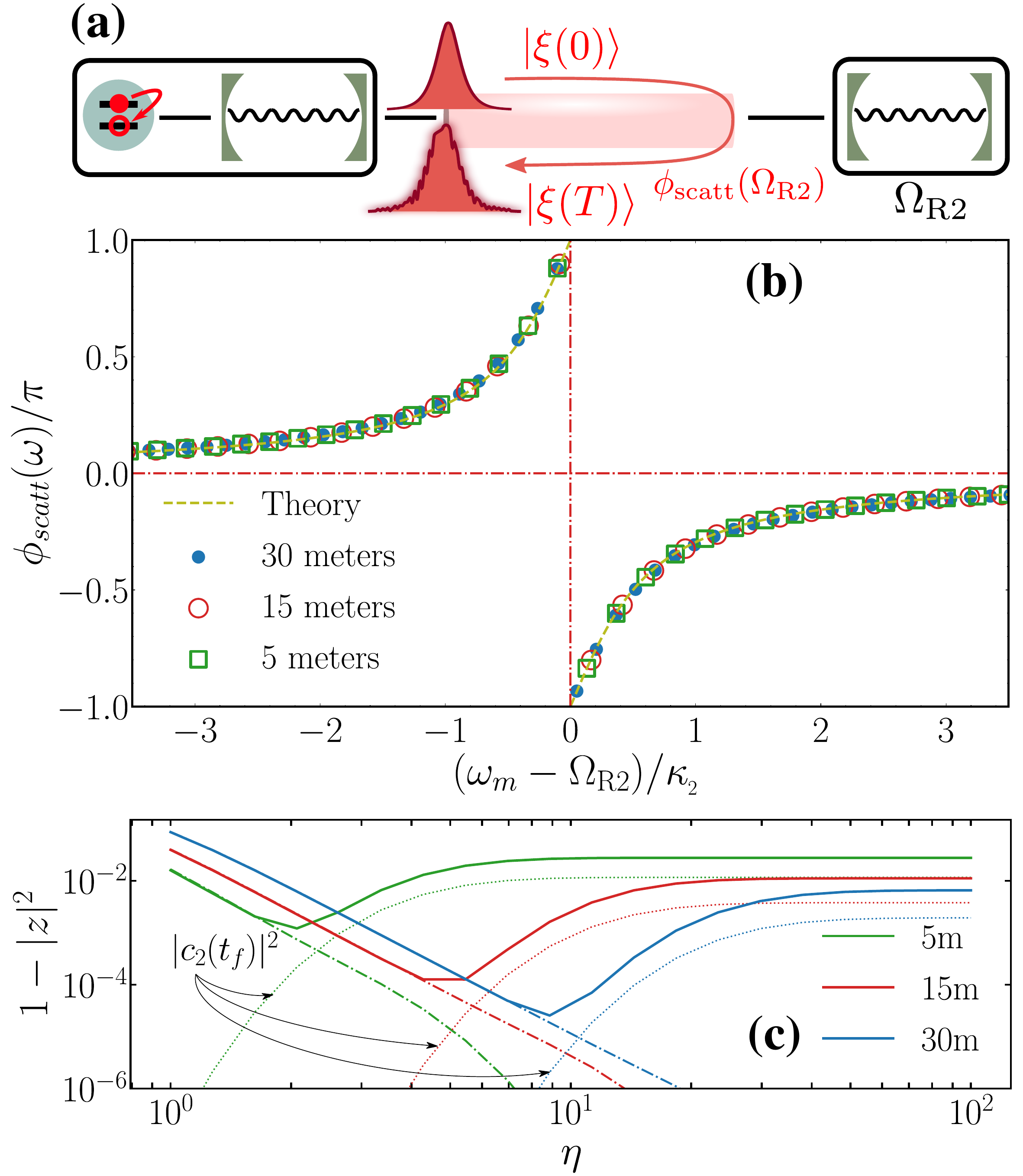}
    \caption{(a) Schematic illustration of the photon scattering process, where the photon interacts with R2, which results in a phase $\phi_{\rm scatt}(\Omega_{\rm R2})$ imprinted on the photon. (b) Phase gained by each mode of frequency $\omega_m$ of the QL after a scattering process with R2 with frequency $\Omega_{\rm R2}$ and decay rate $\kappa_2$. The phase for each of the discrete modes (points) of a QL with $L=30$, $15$ and $5$ meters with $\kappa_2/2\pi=20$, $25$ and $80$ MHz, respectively, closely follows the theoretical expression~\eqref{eq:phiscatt} for any $L$ and $\kappa_2$. Note that the spacing between the points is given by the free spectral range of the QL.
    Panel (c) shows the distortion of a realistic wavepacket due to the phase profile $\phi_{\rm scatt}(\omega)$ as well as due to the residual population that remains in the R2 (dotted lines) for different length as in (b) but with $\kappa=2\pi\times 100$ MHz. Solid lines corresponds to the numerical simulation compared to the ideal photon after scattering $|z|^2 = |\langle{\tilde{\xi}(T)}\ket{\psi(T)}|^2$, while the dashed-dotted lines have been obtained using the theoretical expression~\eqref{eq:phiscatt}, $|z|^2 = |\langle{\tilde{\xi}(T)}\ket{\xi(T)}|^2$. Note that for $\eta\lesssim 10$,  $1-|z|^2\sim \eta^{-4}$ as predicted by Eq.~\eqref{eq:zscatt_theo}, while QLs with larger length $L$ are more prone to distortion due to propagation (cf. Sec.~\ref{sec:QST}). For $\eta\gg 1$ one enters in the long-photon limit where the photon does not fit within the QL and residual populations remain (see main text for further details).}
    \label{fig4}
\end{figure}

According to the standard input-output formalism~\cite{GardinerUltracoldII}, the output field scattered by a resonator R2 with frequency $\Omega_{\rm R2}$ experiences a linear transformation $b_{\rm out}=e^{i\phi_{\rm scatt}(\omega)}b_{\rm in}$ that preserves the population of each mode and only imprints a frequency-dependent phase
\begin{align}\label{eq:phiscatt}
    e^{i\phi_{\rm scatt}(\omega)}=\frac{i(\omega-\Omega_{\rm R2})+\kappa_2/2}{i(\omega-\Omega_{\rm R2})-\kappa_2/2}.
\end{align}
Here, $\omega$ and $\kappa$ are the frequency of the incident photon and decay rate of R2, respectively, and the relation is derived in the infinite waveguide limit.

We have verified numerically that in our setup this relation holds for each of the discrete modes $\omega_m$ of the QL. For that, we prepared monochromatic photons $b^\dagger_{\omega}\ket{\mathbf{0}}$ while keeping the Q1-R1 coupling switched off, and verified after a long enough interaction time $T>t_p$ that these photons remain in the original mode only acquiring a phase shift $e^{i\phi_\text{tot}}b^\dagger_{\omega}\ket{\mathbf{0}}$. Such phase being a sum $\phi_\text{tot}=-\omega t + \phi_\text{scat}(\omega)$ of the mode's free evolution $\propto \omega t$ and the scattering term~\eqref{eq:phiscatt}.

Fig.~\ref{fig4}(b) shows the scattering phase experienced by each of the modes in waveguides from $L=5$m up to 30m, once we subtracted the part due to free evolution $\omega t$. The phase profile matches the theoretical expression~\eqref{eq:phiscatt} to a very good approximation, regardless of the length $L$ of the link and of the decay rate $\kappa_2$ of the second resonator. As in the state transfer situation, this is a validation of the input-output theory in regimes where it is not expected to hold.

\subsection{Finite width wavepackets}

Despite the agreement of $\phi_\text{tot}(\omega)$ with the theoretical model, the total phase experienced is the sum of two contributions---the phase accounting for propagation and the scattering phase---which are not linear in the quasimomentum of the photon. Indeed, as we will see now, the curvature of the scattering phase in Fig.~\ref{fig4}(b) has similar effect to the diffraction of the photon studied in Sec.~\ref{sec:QST}.

Let us consider the distortion of a generic photon, created by the first node using the control~\eqref{eq:gt}, that is $\ket{\xi(t=0)}=\int\!f(\omega) b^\dagger_\omega\ket{{\bf 0}}\mathrm{d}\omega$
with $f(\omega)=\sqrt{\pi/(2\tilde{\kappa})}{\rm sech}(\pi(\omega-\omega_c)/\tilde{\kappa})$ and $\kappa_{1,2}=\kappa$. After a sufficiently long time $T>t_p$, the complete photon is scattered by the second resonator R2, which becomes empty. Following our previous theory, the output photon acquires a frequency-dependent phase shift $\ket{\xi(T)}=\int\!f(\omega) e^{-i T\omega} e^{i\phi_{\rm scatt}(\omega)}b^\dagger_\omega\ket{{\bf 0}}\mathrm{d}\omega$. As explained before, the free evolution $e^{iT\omega}$ already causes the diffraction of the photon along the waveguide. We will now analyze the second phase to understand its effect.

For narrow resonant wavepackets centered around frequency $\omega_c$, we can expand $\phi_{\rm scatt}(\omega)$ up to second order in $(\omega-\omega_c)$, so that
$\phi_{\rm scatt}(\omega)\approx \phi_{\rm scatt}(\omega_c)+\phi'(\omega_c)(\omega-\omega_c)+\phi''(\omega_c)(\omega-\omega_c)^2/2$. The zeroth order accounts for the average phase experienced by the photon in the limit of near-zero bandwidth. The first order  $\phi'_{\rm scatt}(\omega)=4\kappa/(\kappa^2+4(\Omega_{\rm R2}-\omega)^2)$ can be interpreted as a delay caused by the absorption and reemission of the photon in R2. Hence, a round trip of the scattered photon takes a time $T=2t_p+\phi'(\omega_c)$ with $t_p$ the propagation time $t_p=L/v_g$. Finally, the quadratic term  accounts for the curvature of the phase profile, and will be responsible for photon distortion at leading order, $\phi''_{\rm scatt}(\omega)=32\kappa(\Omega_{\rm R2}-\omega)/(\kappa^2+4(\Omega_{\rm R2}-\omega)^2)^2$.

The real scattered photon $\ket{\xi(T)}$ can be compared with an ideal photon $\ket{\tilde{\xi}(T)}$ that experiences no distortion, is only delayed and acquires an average scattering phase $\bar{\phi}_\text{scat}\simeq \phi_\text{scat}(\omega_c)$. We can quantify the fidelity of the process by analyzing the overlap between both wavepackets, in a formula that includes the curvate of the phase profile and propagation through the waveguide (cf. App.~\ref{app:dis})
\begin{align}\label{eq:zscatt_theo}
    |z|^2&=\left|\langle \tilde{\xi}(T) | \xi(T) \rangle \right|^2=\left| \int d\omega \ |f(\omega)|^2 e^{i(\phi''_{\rm scatt}(\omega_c)-T D/v_g^2)(\omega-\omega_c)^2/2}\right|^2\nonumber\\
    &\approx 1-\frac{\kappa^4}{45\eta^4}\left(\frac{\phi''_{\rm scatt}(\omega_c)}{2}-\frac{T D}{2v_g^2}\right)^2,
\end{align}
and also by studying how the resulting phase of the scattered photon deviates from the prediction (cf. App.~\ref{app:dis}).

We have compared our theoretical predictions with an exact numerical simulation of the scattering dynamics, for a given frequency of the resonator. Below, in Sec.~\ref{ss:cphase} , this will be controlled by the state of the qubit, but for now, we arbitrarily set $\omega_c-\Omega_{\rm R2}=\kappa(1/2+1/\sqrt{2})$ so that the resulting phase gained by the photon upon scattering is $\phi_{\rm scatt}(\omega_c)=-\pi/4$ (cf. Fig.~\ref{fig4}(b)), while $\phi''_{\rm scatt}(\omega_c)=2(1-\sqrt{2})/\kappa^2$.

We compute $|z|^2$ in two different ways in order to compare our analytical second-order perturbation theory with the simulation. First, we calculate the overlap between the ideal non-distorted wavepacket $|\tilde{\xi}(T)\rangle$ and the numerically simulated one $\ket{\psi(T)}=e^{-iT (H_{\rm QL}+H_{\rm N_2-QL})}\ket{\psi(0)}$ where $\ket{\psi(0)}$ denotes the initial photon in the QL and empty R2 ($c_2(0)=0$). And second, we take the same non-distorted $|\tilde{\xi}(T)\rangle$ and project it onto the theoretically predicted one, $\ket{\xi(T)}=\sum_m e^{-iT\omega_m}e^{i\phi_{\rm scatt}(\omega_m)}f(\omega_m)b^\dagger_m\ket{{\bf 0}}$, with $\phi_{\rm scatt}(\omega)$ given by Eq.~\eqref{eq:phiscatt}.

As we illustrate in Fig.~\ref{fig4}(c), in order to achieve a low distortion and good matching of the scattering phase, the scattered photons must have a bandwidth $\tilde\kappa$ that is narrower than $\kappa_2$ of the second resonator. In this limit, the infidelities follow closely the theoretical predictions and decrease monotonically with the factor $\eta=\kappa_2/\tilde\kappa$, up to a minimum bandwidth that depends on the length $L$ of the waveguide.

At this point, the infidelity begins to grow again due to the transition to the long-photon limit. More precisely, as $\eta$ increases beyond this point, the photon gains a larger temporal width $\sigma_t=\pi\eta/(\sqrt{3}\kappa)$, which becomes longer than the propagation time between nodes $\sigma_t\gtrsim t_p$. In this limit the photon no longer fits within the waveguide and the dynamics changes radically, because the resonator R2 is not capable of becoming completely empty, and acquires a residual population even long after the collision $T=2t_p+\phi_{\rm scatt}(\omega_c)$. This residual population accounts for a rapid decrease in the norm of the scattered wavepacket and the associated growth of the infidelity.

Summing up, the interaction between a propagating photon created by a node Q1-R1 and a second resonator R2 at a different node is very well described by input-output theory. Unlike the original proposal~\cite{Duan2004}, we have found that it is very important to consider the shape and duration of the wavepacket and the separation between the nodes. In practical applications the long-photon limit will be of little practical relevance, because we will need photons of intermediate size  ($\sigma_t\lesssim t_p$), where we can physically distinguish the stages of photon emission and scattering, from the reabsorption of the photon by the Q1-R1 node (cf. Sec.~\ref{ss:cphase}).

\section{Applications}
\label{sec:applications}

\subsection{Quantum links}

Quantum links have been put forward as a means to distribute quantum information and implement quantum protocols among separate quantum nodes. According to what we have seen in Sec.~\ref{sec:QST}, the same pulse shaping protocols can be used to implement realistic quantum links of almost any size, as intra- or inter-node quantum links, with fidelities that are very competitive even if one accounts for diffraction-induced errors. In particular, as we have seen, quantum links may be created using commercial waveguides, without circulators~\cite{Magnard2020}, and with faster rates than in adiabatic protocols~\cite{Chang2020}.

Based on this, we could take the state transfer primitive as discussed in this work and use it to build a distributed quantum computing toolbox. As it is usual in the literature, operations in this toolbox would require four steps: (i) creating entangling pairs in different nodes, (ii) distributing this entanglement with state transfer, (iii) distilling this entanglement if the quality is insufficient, (iv) implementing quantum operations via quantum teleportation or similar protocols.

This scheme is not very well suited for practical application in current superconducting architectures, because the implementation of the gates requires measurements and classical communications, between separate nodes. The measurement introduces important errors, of about $1\%$ in the fidelity and even greater if we consider the distortion of the post-measurement state. The classical communication requires sophisticated synchronization and slows down the gate, increasing the probability that it is affected by decoherence.

Motivated by this, we now introduce two alternative implementations of a truly distributed quantum gate among nodes connected by a quantum link. The first approach replaces the teleportation-based protocol---with one round of quantum transfer and one of classical communication---with a protocol where the bidirectional communication is fully quantum. The second approach is a passive protocol, which does not require synchronization between the nodes, where the gate is implemented by means of scattering.

\subsection{Quantum gate transfer}\label{ss:QGT}

\begin{figure}
    \includegraphics[width=\columnwidth]{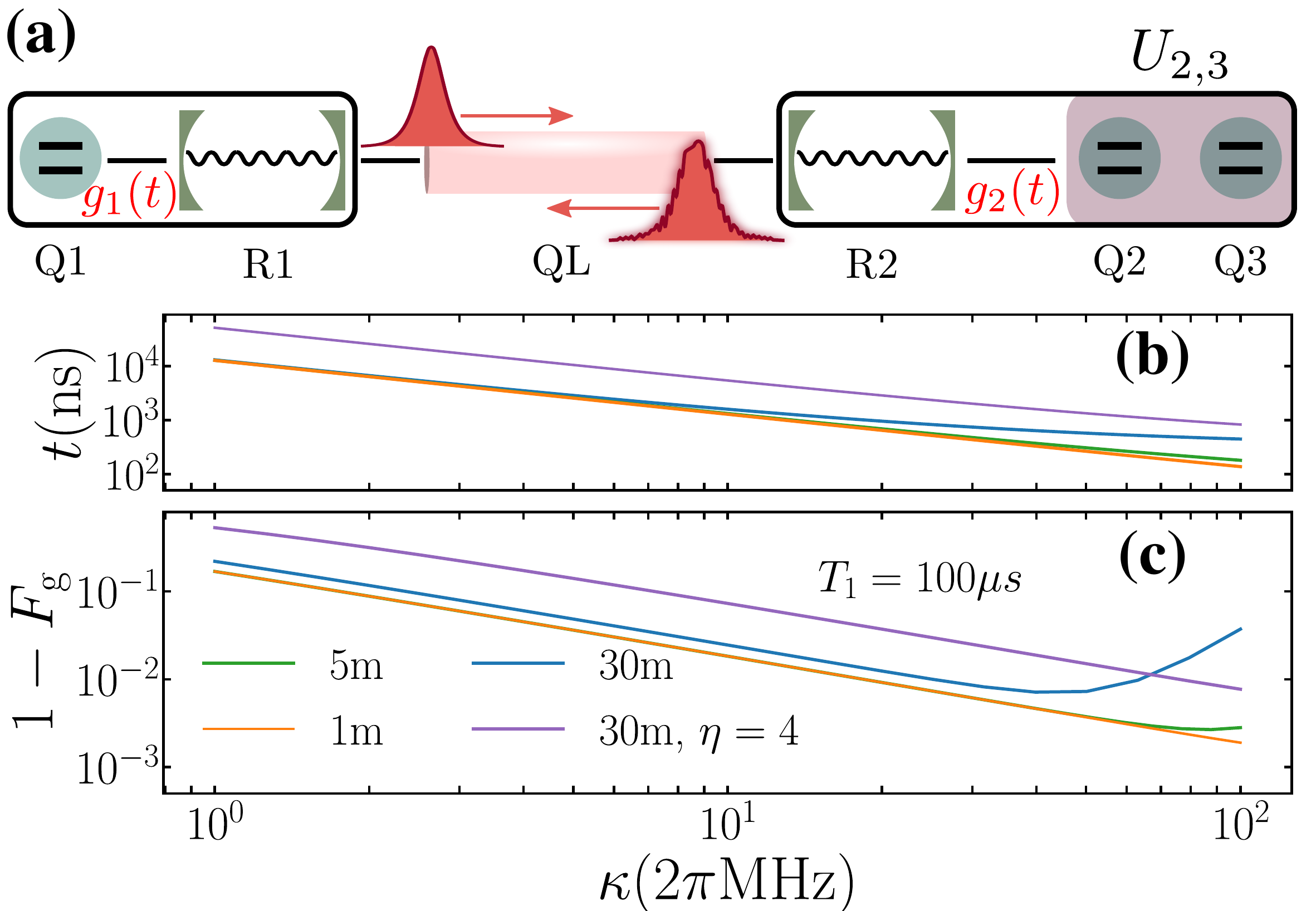}
    \caption{(a) Schematic illustration of a quantum gate transfer between Q1 and Q3, as given in Eq.~\eqref{eq:QGT}, that involves two state transfer operations and a two-qubit gate $U_{2,3}$ on qubits within the same node. (b) Duration of the gate transfer protocol $t = 2\times 2T$, assuming that the local gate implemented at the second node is instantaneous, for three different waveguide lengths with $\eta = 1$ (omitted in the legend) and one case with $L=30$ m and $\eta=4$. For small $\kappa$ the propagation time is negligible, and the limitation comes from the time-dependent control $g(t)$ and all lines with $\eta=1$ overlap. (c) Fidelity of the process for the same set of parameters, where a trade-off between protocol duration and distortion of the wave packet is observed, specially in the $L=30$ m, $\eta = 1$ case.}
    \label{fig5}
\end{figure}

The \emph{quantum gate transfer} is a protocol that uses state transfer to distribute a quantum operation, much like we would distribute an entangled state. The gate transfer
\begin{equation}\label{eq:QGT}
  U_{1,3}\otimes\openone_2 := \mathcal{T}_{1,2}^\dagger U_{2,3} \mathcal{T}_{1,2}
\end{equation}
consists of two state transfer protocols with opposite directions---$\mathcal{T}_{1,2}$ from Eq.~\eqref{eq:ST} and its inverse $\mathcal{T}_{2,1}$---surrounding a two-qubit gate $U_{2,3}$ that takes place between qubits Q2 and Q3 in the second node (cf. Fig.~\ref{fig5}(a)).

Qualitatively, the second state transfer protocol replaces the classical communication that would be required for a quantum teleportation-based gate. Even though this communication happens at a fraction of the speed of light $2c/3$, the concatenated operation may perform better than the quantum-classical protocol. This is true, in particular for platforms such as superconducting circuits, where a state transfer can be implemented with high fidelity while the overhead for quantum state distillation is large.

We can lower bound the fidelity of the gate transfer
\begin{equation}
  F=\left[F_\text{st}\exp(-\tau/T_1)\right]^2 F_{2,3},
\end{equation}
based on the fidelity $F_{2,3}$ of the operation in a single node $U_{2,3}$, the fidelity of a single step of state transfer $F_\text{st}=|q_2(T)|^2$, the single qubit decoherence time $T_1$ (or any other decoherence rates that can be accounted for, such as non-radiative decay), and where $\tau$ is the time during which the qubits are populated in one quantum state transfer (cf. Sec.~\ref{ss:imp}).

We can analyze this estimate, ignoring the constant factor $F_{2,3}$, and using the values of $F_\text{st}$ and $T$ from the simulations in Sec.~\ref{sec:QST}. As shown in Fig.~\ref{fig5}(b), the quantum gate transfer has an optimal operation point which is a compromise between distortion errors and decoherence induced by the protocol duration. We find realistic scenarios with competitive gate fidelities $1-F\sim 10^{-2}-10^{-3}$, specially for the short waveguides that would be required to connect nearby fridges, or chips within the same fridge. It is worth recalling  that the protocols that we have employed allow for a faster control than adiabatic transfer of population, as it was made explicit in Sec.~\ref{sec:QST} (cf. Fig.~\ref{fig3}).

We can further optimize the fidelity of the gate transfer by reducing the protocol duration $2T$. The data shown in Figs.~\ref{fig5}(b)-(c) was obtained by imposing that each point saturates to the distortion by propagation limit. As already commented, this is a requirement that might be too restrictive for real-world applications.  As an example to illustrate this fact, we have optimized the protocol duration for the specific set of parameters $L=30$ m, $\kappa=2\pi\times 3$ MHz, $T_1 = 100\ \mu$s and have obtained a combined gate fidelity of $F_g \approx 10^{-2}$, a $5-$fold improvement with respect to the results with a naive choice of $2T$.

\subsection{Controlled-phase gate}\label{ss:cphase}

\begin{figure}
    \includegraphics[width=\columnwidth]{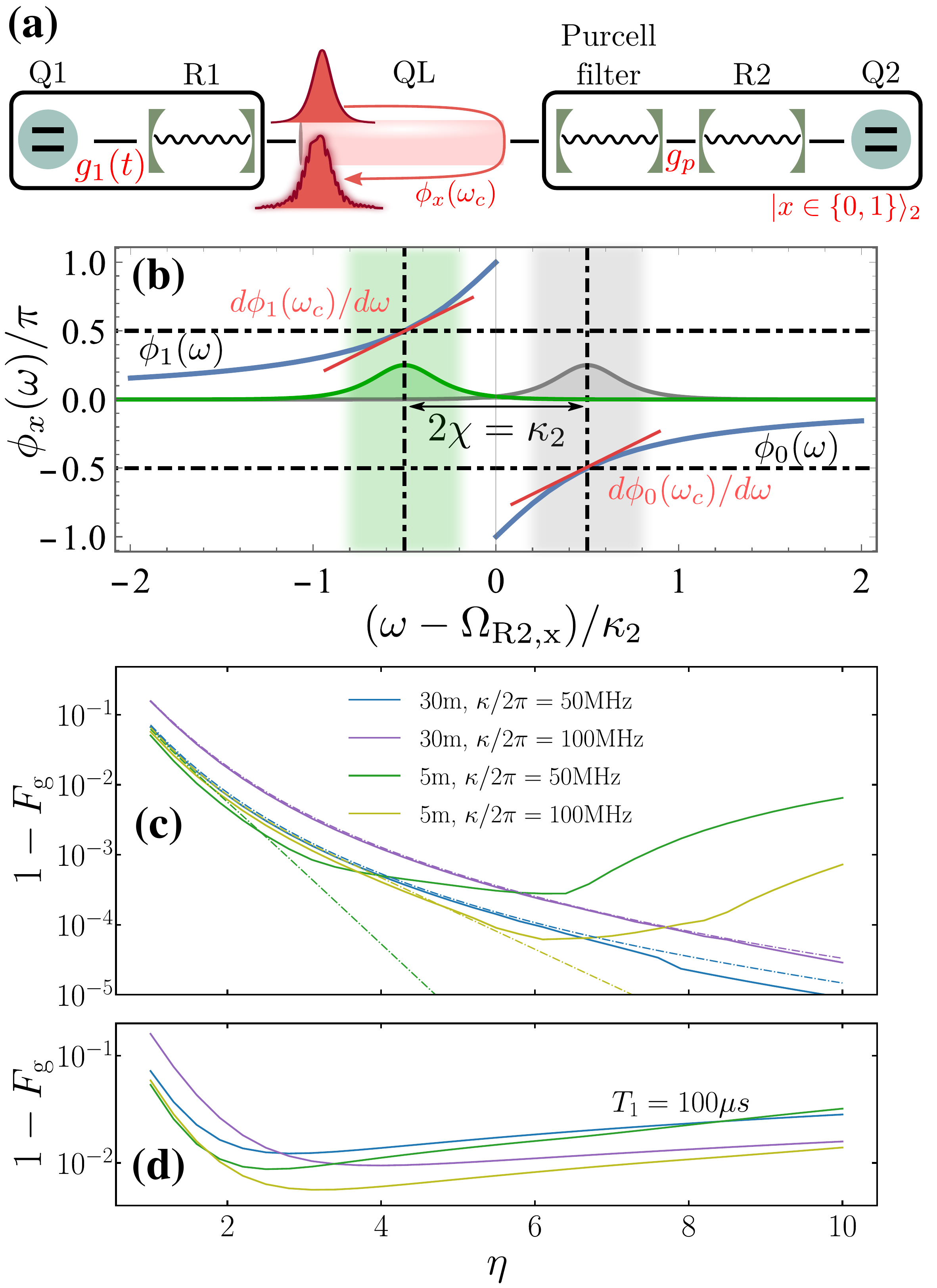}
\caption{\small{(a) Schematic illustration for the realization of a controlled-phase gate, where the phase of the scattered photon depends on the Q2's state, $\ket{x\in\{1,0\}}$. (b) Phase profile $\phi_x(\omega)$ as a function of the detuning of the incident photon with respect to the effective R2 frequency. (c) Infidelity of the controlled-phase gate $1-F_g$ for two different $L$ and $\kappa$. The solid lines correspond to the numerical simulations optimizing $\chi$ so one can access to the long-photon regime, while the theoretical predictions (dashed lines) are calculated from the joint diffraction and scattering expressions. Nor decoherence times nor the effect of the Purcell filter are considered. (d) Results for $1-F_g$ for the same set of parameters but including coherence times and a Purcell filter between QL and R2. We set $g_2/\Delta=0.1$ and $0.125$ for $\kappa=2\pi \times 100$ and $50$ MHz with $g_p/\Delta=0.03$ and $0.04$, respectively.}}
\label{fig6}
\end{figure}

The gate transfer operation requires perfect synchronization between the emitting and absorbing nodes during state transfer. We can engineer a simpler gate, in which only the first node Q1-R1 emits and reabsorbs a photon that gets entangled with the second node, using the photon-cavity scattering phases from Ref.~\cite{Duan2004} and Sec.~\ref{sec:scatt}. In this protocol only the first node requires an explicit control, and the second node only enters as a resonance frequency $\Omega_\text{R2}+\chi\sigma^z_2$ whose location shifts depending on the state of the second qubit (Q2 in Fig.~\ref{fig6}(a)).

Qualitatively, we will map the state of qubit Q1 to the absence or the presence of travelling photon $\ket{\xi_\text{in}}$. This photon travels through the link and is scattered by the second node, acquiring a qubit-dependent scattering phase $\ket{\xi_{\rm in}} \left(\alpha \ket{0}_2+\beta \ket{1}_2\right)
\longrightarrow \left( \ket{\xi_{\rm out}} \ket{0}_2 \alpha e^{-i\varphi_0} +\ket{\xi_{\rm out}} \ket{1}_2 \beta e^{-i\varphi_1  } \right)$. The travelling photon $\ket{\xi_\text{out}}$ is then reabsorbed by the first node using the same tools of state transfer. Tuning the resonance's center $\Omega_\text{R2}$ and displacement $\chi$, we engineer a condition $\varphi=\varphi_1-\varphi_0=\pi$ such that, once the photon is reabsorbed, the result is a controlled-phase gate between Q1 and Q2
\begin{align}\label{eq:cphase}
U_{\rm cp} = & \text{exp}\left[-i \frac{1}{2}(\sigma^z_1+1)\otimes \frac{\pi}{2}\sigma^z_2\right].
\end{align}

Let us first analyze the experimental configuration from Ref.~\cite{Duan2004} and Fig.~\ref{fig1}(b), in which the second node only has a single cavity, whose frequency is dispersively shifted by a qubit. In the superconducting qubit scenario, this coupling is described by the Hamiltonian~\cite{Kono2018,Besse2018}
\begin{equation}\label{eq:H2}
  H_{\rm N_2}=\delta_2 \sigma^+_2\sigma_2^-+(\Omega_{\rm R2} +\chi \sigma^z_2) a^\dagger_2 a_2.
\end{equation}
Depending on Q2's state $\ket{x\in\{0,1\}}$, the resonator R2 experiences a frequency shift $\Omega_{\rm R2,x}=\Omega_{\rm R2}+(-1)^{x+1}\chi$, where $\chi$ is a function of the Q2-R2 coupling $g_2$ and their detuning $\Delta = \delta_2-\Omega_{\rm R2}$. The dispersive limit~\eqref{eq:H2} is valid as long as $\Delta \gg g_2$, which is routinely fulfilled in many experiments.

As described by Ref.~\cite{Duan2004} and analyzed in Sec.~\ref{sec:scatt}, a quasi-monochromatic photon scattered by R2 acquires a phase shift $b^\dagger_\omega\ket{{\bf 0}}\ket{x} \rightarrow e^{i\phi(\omega,x)}b^\dagger_\omega\ket{{\bf 0}}\ket{x}$ with the phase $\phi(\omega,x)$ given in Eq.~\eqref{eq:phiscatt} with $\Omega_{\rm R2,x}$.
At resonance $\omega_c = \Omega_{\rm R2}$ with $\chi = \kappa_2/2$, the phase difference between qubit states is maximal $\phi(\omega_c,x)=(-1)^{x+1}\pi/2$, and we reach the ideal gate $\varphi_{\rm id}=\pi$ (cf. Fig.~\ref{fig6}(b)).

In practice, as discussed in Sec.~\ref{sec:scatt}, the first node can only create photons with a finite bandwidth. These photons are prone to diffraction induced by the waveguide and the curvature of the scattering profile. These imperfections can be estimated as (cf. App.~\ref{app:fid})
\begin{align}\label{eq:zx_cphase}
    |z_x|^2&\approx 1-\frac{\kappa_1^4}{45\eta^4}\left(\frac{ t_p D}{v_g^2} + (-1)^x\frac{2}{\kappa_2^2}\right)^2.
\end{align}
This formula reveals an asymmetric behavior $|z_1|\geq |z_0|$, whereby for $x=1$ the distortions caused by scattering and propagation compensate each other, while for $x=0$ both contributions add up. From the value of $z_x$ and the phase difference $\varphi$ we obtain a lower-bound to the gate fidelity, i.e. the minimum gate fidelity~\cite{Nielsen}, with respect to the ideal gate operation $U_\text{cp}$. Following App.~\ref{app:fid}, the bound is
\begin{align}\label{eq:fidelity}
    F_g=\begin{cases} \frac{|z_0|^2\sin^2\varphi}{(1+r^2+2r\cos\varphi)}, \quad &{\rm if} \quad \frac{ (1+r\cos\varphi)}{(1+r^2+2r\cos\varphi)}\leq 1,\\ |z_0|^2, \quad &{\rm otherwise}, \end{cases}
\end{align}
where we have defined $r \equiv |z_0|/|z_1| \leq 1$.

We have simulated numerically the gate operation for a setup with $\kappa_{1,2}=\kappa$. The passive node Q2-R2 has a constant coupling $g_2(t)=g_2$ that engineers the dispersive interaction~\eqref{eq:H2}. The node Q1-R1 is controlled with a symmetric pulse for emission and reabsorption of a photon of width $\tilde{\kappa}\leq\kappa$
\begin{equation}\label{eq:reabsorption}
  g_1(t) = \begin{cases} g(t+t_d/2;\tilde\kappa,\kappa), & \mbox{for } t\in[-T,0],\\
    g(-t+t_d/2;\tilde\kappa,\kappa), & t\in[0,T].
  \end{cases}
\end{equation}
The delay $t_d= 2L/v_g + 2/\kappa$ is now given by the photon's propagation and the delay intrinsic to scattering (cf. Sec.~\ref{sec:scatt}).

As shown in Fig.~\ref{fig6}(c), the gate fidelity adjusts well to the theoretical model. Diffraction makes the fidelity decrease both with the waveguide length, and with the ratio $\eta=\kappa/\tilde\kappa$ of the photon's bandwidth to the scattering phase curvature. The gate fidelity reaches very good values for moderate $\eta$, at the expense of protocol duration. As discussed before, the gate still works beyond a limit where the input-output approximation breaks down. However, as system enters in the long-photon regime---$\eta\gtrsim 2$ (5) for $L=5$m (10m)---the optimal value of $\chi$ deviates from $\chi=\kappa/2$ and must be computed semi-analytically. Also, in the regime where the photon is wider than twice the quantum link there is an overlap between the emission and absorption controls which makes it impossible for perfect reabsorption, as it can be derived from~\eqref{eq:reabsorption}.

This setup allows us to study the fundamental limitations of the gate induced by the scattering processes. However, the dispersive coupling between the qubit and the cavity---at least in the usual models for superconducting qubits---introduces a Purcell-induced decay in the data qubit Q2, $\gamma_\kappa=\kappa_2(g_2/\Delta)^2$~\cite{Boissonneault2009, Jeffrey2014}. This decay process has not been taken into account in the discussion above, but it can significantly degrade the gate fidelity. This effect can be mitigated by the introduction of a Purcell filter~\cite{Sete2015,Walter2017,Kurpiers2017}, which can enhance the coherence of the  dispersively-coupled qubit by two orders of magnitude. The Purcell filter is another resonator, placed in between R2 and the QL (cf. Fig.~\ref{fig6}(a)), which effectively decreases the probability that Q2 decays to the line.

We have considered a resonant cavity coupled with decay rate $\kappa_2$ to the QL and with strength $g_p$ to R2.
With the introduction of the filter the Q2's decay rate is now reduced by $\gamma_\kappa\approx\kappa_2(g_2/\Delta)^2(g_p/\Delta)^2$ ~\cite{Sete2015}, given that $\Delta\gg g_2,G_2,\kappa_2$. Furthermore, the equation for the phase profile \eqref{eq:cphase} still holds and the overall gate fidelity results from multiplying Fig.~\ref{fig6}(c)  by a factor  $e^{-(\tau/T_1 + T\gamma_\kappa)}$. For the unoptimized parameters in Fig.~\ref{fig6}(d), we find $1/\gamma_\kappa\gtrsim T_1$ with $T_1=100\mu$s. Under these realistic conditions, we observe that a controlled-phase gate can be implemented with a fidelity $1-F_g\approx 10^{-2}$.

\section{Conclusions and outlook}
\label{sec:summary}

In this article we have designed protocols that allow for the realization of universal two-qubit quantum gates in a distributed and deterministic fashion between quantum nodes within a network connected via quantum links. These protocols build on the primitives of quantum state transfer~\cite{Cirac1996} and photon-cavity entangling gates~\cite{Duan2004}, whose performance we analyze through a realistic modelization of state-of-the-art hardware~\cite{Kurpiers2017}, including non-linear dispersion relations, finite length links, finite-bandwidth photons and current qubit coherence times.

From a physical point of view, our numerical simulation of the device reveals a surprising picture in which the input-output theory used to derive both primitives~\cite{Cirac1996,Duan2004} still works in limits of large FSR and short links. In particular, we show that the state transfer controls based on pulse shaping  operate even in regimes where they are not expected to hold, that is, for short length and/or few relevant modes within the quantum link, while allowing for fast operation limited by the cavity bandwidth. Indeed, pulse shaping outperforms standard adiabatic protocols for quantum state transfer, even for short-length quantum links.
Considering realistic coherence times $T_1=20-100\ \mu$s, we find competitive gate fidelities $1-F=10^{-2}-10^{-3}$. We believe that these gates will be comparable or significantly better to other alternatives based on entanglement distribution, specially when one considers the errors induced by measurements and by slow classical communication.

In light of the reported results, we believe that quantum networks made of superconducting qubits and wavguides constitute a promising platform to implement  quantum operations on both long-distant (inter-node) and short-distance (intra-node) quantum processors. Furthermore, we expect further theoretical developments, such as chirping~\cite{Casulleras2021} and quantum control \cite{Stannigel2010, Stannigel2011}, to overcome the limitations induced by propagation and Stark shifts, and to further optimize the speed of state transfer. Finally, all the codes that support the simulations of this work at available at \cite{gphysics_fpf_2022_6025312}.

\begin{acknowledgements}
 This work has been supported by the European Union's Horizon 2020 FET-Open project SuperQuLAN (899354) and Comunidad de Madrid Sin\'ergicos 2020 project NanoQuCo-CM (Y2020/TCS-6545). We acknowledge valuable feedback given by Anatoly Kulikov, Simon Storz and Josua Sch{\"a}r.  T.R. further acknowledges support from the EU Horizon 2020 program under the Marie Sk\l{}odowska-Curie grant agreement No. 798397, and from the Juan de la Cierva fellowship IJC2019-040260-I.
\end{acknowledgements}

\appendix

\section{Pulse shaping}\label{app:pulse}
Here we provide the details to derive Eq.~\eqref{eq:gt}. For that, we consider a qubit interacting with the resonator, which in turns decays at rate $\kappa$ into the QL. This minimal model can be written as
\begin{align}
    \dot{q}(t)&=-i g(t) c(t)\\
    \dot{c}(t)&=-i g(t) q(t)-\frac{\kappa}{2}c(t),
\end{align}
where $g(t)\in\mathbb{R}$, and $q(t)$ and $c(t)$ denote the amplitude of the state with one excitation in the qubit and resonator, respectively. Defining $\tilde{c}(t)=-ic(t)$, so that $q(t),\tilde{c}(t)\in\mathbb{R}$, the outgoing photon is given by $\psi(t)=\sqrt{\kappa}\tilde{c}(t)$.
The probability that the excitation remains in the qubit or resonator is $d(q^2(t) +\tilde{c}^2(t))/dt=-\kappa \tilde{c}^2(t)$, which can be solved formally as
\begin{align}
    q^2(t)=q^2(t_0)-\frac{1}{\kappa}\psi^2(t)-\int_{t_0}^{t}d\tau \ \psi^2(\tau).
\end{align}
This expression sets a constraint on the pulse evolution, $0\leq \frac{1}{\kappa}\psi^2(t)+\int_{t_0}^{t}d\tau \ \psi^2(\tau)\leq 1$. Once we solve $q^2(t)$ we can obtain the pulse as $g(t)=\sqrt{\kappa}\dot{q}(t)/\psi(t)$ imposing the form of the photon as $\psi(t)=-\sqrt{\tilde{\kappa}/4}{\rm sech}(\tilde{\kappa}t/2)$,
so that $\int_{-\infty}^{\infty} dt |\psi(t)|^2=1$, whose Fourier transformed gives $f(\omega)$. Following the previous steps, we obtain $g(t)$ given in Eq.~\eqref{eq:gt}, with the condition $\tilde{\kappa}\leq \kappa$.

\section{Control distortion due to a limited bandwidth}\label{app:Bandwidth}
The control $g(t)$ with which we tune the coupling between qubit and resonator (cf. Eq. \ref{eq:gt}) is a smooth function with low-frequency components, and therefore, it is robust against a limited bandwidth, as we show below.

Although a large $\tilde{\kappa}$ allows for a faster operation, $g(t)$ becomes wider in frequency space. In this manner,  a limited bandwidth may distort the control.  One natural way to quantify this effect is to consider the impact that adding a low-pass filter has on the control $g(t)$~\cite{Langford2017,Ripoll2020}.

\begin{figure}
    \includegraphics[width=\columnwidth]{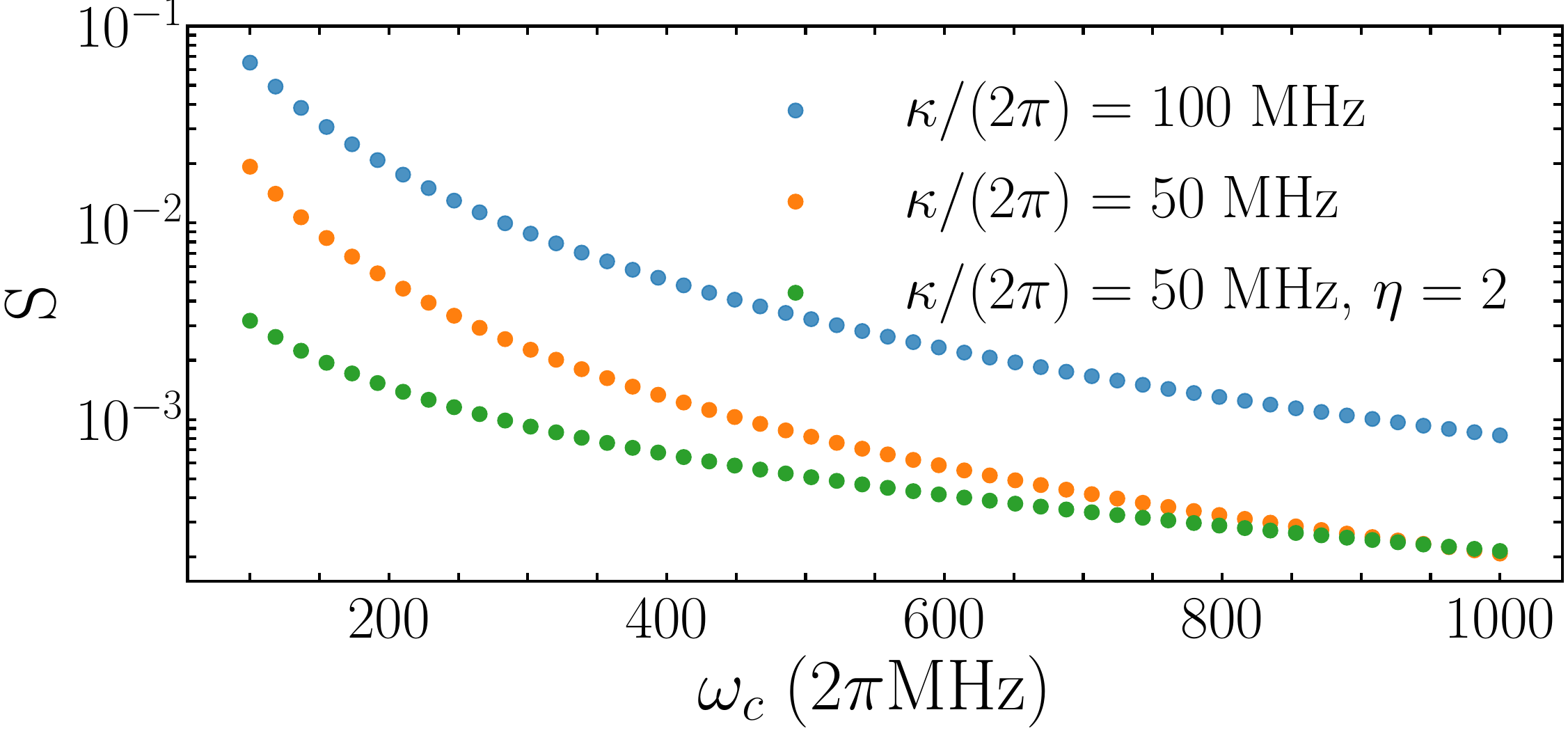}
\caption{Dependence of the distortion $S$ (cf. Eq.~\eqref{eq:controldist}) in the control $g(t)$ (cf. Eq.~\eqref{eq:gt}) produced by a low-pass filter with cut-off frequency $\omega_c$. Recall that $\kappa$ is the resonator decay rate.}
\label{fig7}
\end{figure}

Considering the control $g(t)$ given in Eq.~\eqref{eq:gt}, the actual output signal that is injected into the system is given by a filtered function $\tilde{g}(t)$, such that $\tilde{g}(t) = \mathcal{F}^{-1} \left[ \mathcal{F}[g(t)] \mathcal{H(\omega)} \right]$, where $\mathcal{H}(\omega)$ is the low-pass filter function and $\mathcal{F}[\cdot]$ the Fourier transform. Thus, we quantify the distortion due to a limited bandwidth limitations by comparing $\tilde{g}(t)$ with the ideal control function $g(t)$. For this purpose we define the following quantity
\begin{equation}\label{eq:controldist}
S= 1 - \frac{\int dt \ |g(t)| |\tilde{g}(t)| }{\int dt \ |g(t)|^2 }.\end{equation}
Following~\cite{Langford2017}, we consider a transfer function $\mathcal{H(\omega)} = \omega_c/(\omega_c-i\omega)$ that captures the effect of placing a low-pass filter with cutoff frequency $\omega_c$. In Fig.~\ref{fig7} we show the value of the distortion $S$ as a function of the cut-off frequency of the filter for three different set of parameters. Note that for $\omega_c/\kappa \approx 10$, which is a regime easily accessible in current experiments, $S\approx 10^{-3}$. 

\section{Quantification of photon distortion}\label{app:dis}

The overlap between a distorted photon $|\xi(t)\rangle$ and without distortion $|\tilde{\xi}(t)\rangle$, can be approximated by
\begin{align}\label{eq:Flimit}
    |z|^2=\left|\langle\tilde{\xi}(t) |\xi(t) \rangle \right|^2=\left|\int d\omega |f(\omega)|^2 e^{-i h (\omega-\omega_c)^2} \right|^2,
\end{align}
where $h$ is just a parameter that depends on the origin of the distortion. The first four moments $m_{n}=\int d\omega\ |f(\omega)|^2 (\omega-\omega_c)^n$ with $f(\omega)=\sqrt{\pi/(2\tilde{\kappa})}{\rm sech}(\pi(\omega-\omega_c)/\tilde{\kappa})$ the photon generated under Eq.~\eqref{eq:gt}, we find are $m_2=\tilde{\kappa}^2/12$ and $m_4=7\tilde{\kappa}^4/240$.  $m_{2n+1}=0$ for $n\geq 0$ due to parity symmetry. 
By further approximating the integral up to fourth order in $(\omega-\omega_c)$, we find $z\approx 1-ih m_2-h^2 m_4/2$, so that
\begin{align}\label{eq:zdist}
|z|^2\approx 1-\frac{h^2\tilde{\kappa}^4}{45}+O(h^4\kappa^8).
\end{align}

Now, for the case of propagation, we can expand the dispersion relation up to second order in the vicinity of the central frequency $\omega_c$,
\begin{align}\label{eq:Omegaklin}
    \omega(k)&\approx
    v_g k+ \frac{(\omega-\omega_c)^2}{2v_g^2}D,
\end{align}
where $D={\rm d}^2\omega(k)/{\rm d}k^2|_{k_c}$. We find $|z|^2$ given in Eq.~\eqref{eq:stfidelity} from Eq.~\eqref{eq:zdist} setting $h=t_p D /2v_g^2$.

In a similar manner, we can expand the phase after scattering $\phi_{\rm scatt}(\omega)$, cf. Eq.~\eqref{eq:phiscatt}, up to second order in $(\omega-\omega_c)$ to tackle the distortion in the scattering. Since the first order can be absorbed as a time delay of the propagation, the main source of imperfection stems from the second order term, which can be quantified again resorting to Eq.~\eqref{eq:zdist} setting $h=TD/2v_g^2- d^2\phi_{\rm scatt}(\omega)/d\omega^2\left.\right|_{\omega_c}/2$. When substituting the corresponding value of the second derivative in~\eqref{eq:zdist}, one immediately finds Eqs.~\eqref{eq:zscatt_theo} and~\eqref{eq:zx_cphase}.

\section{Long-photon limit and dressed modes}\label{app:hyb}
In the long photon limit, one can gain insight studying the R1-QL-R2 dressed states. A diagonalization of their Hamiltonian leads to
\begin{align}
    \Lambda^\dagger H_{\rm R1-QL-R2}\Lambda = \sum_{n}\tilde{\omega}_n c_n^\dagger c_n
\end{align}
where $c_n$ refers to the dressed R1-QL-R2 modes. The central mode $\tilde{\omega}_0$ and the two closest modes to it, $\tilde{\omega}_{\pm 1}$ which have the largest resonator content, and are close to the resonant condition, $\tilde{\omega}_0,\tilde{\omega}_{\pm 1}\approx \delta_{1,2}$.
In an ideal waveguide (linear dispersion relation and assuming constant coupling $G_{m,j}\equiv G$) one can show~\cite{Vogell2017} that
    $\tilde{\omega}^{\rm id}_0=\Omega_{\rm R1,R2}$, and $\tilde{\omega}^{\rm id}_{\pm 1}=\tilde{\omega}^{\rm id}_0 \pm \sqrt{2} G$, assuming $\Omega_{\rm R1}=\Omega_{\rm R2}$.  The small frequency separation between the two adjacent states makes them relevant for the dynamics~\cite{Vogell2017}. A realistic dispersion relation modifies these expressions by an amount of the order of the Lamb shift found numerically, while inducing an asymmetry in the resonator content in the dressed modes.

Including now the qubits, we can write the Hamiltonian as
\begin{align}
    H=\sum_n \tilde{\omega}_n c_n^\dagger c_n +\sum_{j=1,2}\delta_{j}\sigma^+_j\sigma^-_j + g_{j}(t) \sigma_j^+ \left(\sum_n \beta_n^j c_n\right)+{\rm H.c.}\nonumber,
\end{align}
where the $a^\dagger_j$ has been replaced by its transformed expression in the dressed mode picture using the unitary matrix $\Lambda$, $a^\dagger_j\rightarrow \sum_n \beta_n^j c_n$. Proceeding as described in the main text relying on the Wigner-Weisskopf Ansatz, $\ket{\psi(t)}=\left[\sum_{i=1,2}\sigma^+_i q_i(t)+\sum_{n}c_n^\dagger d_n(t) \right]\ket{{\bf 0}}$, we find that $|q_{1}(t)|^2+|q_2(t)|^2+|d_0(t)|^2\approx 1$ in the long photon limit. Yet, there is still a small contribution given by the two adjacent dressed modes $\tilde{\omega}_{\pm 1}$~\cite{Vogell2017}.

\subsection{Stark shifts in the long photon limit}
In the case in which the system reduces to three or few dressed modes, the states may be adiabatically eliminated, which in turn introduce a small time-dependent frequency detuning in the qubits. This can be justified imposing $\dot{d}_m(t)=0$ that leads to a time-dependent Stark shift in the qubits' frequencies,
    $\tilde{\delta}_j(t)=\delta_i-\sum_{m} \frac{g_i^2(t)|\beta_m^i|^2}{\tilde{\omega}_m}$,
and a flip-flop term between qubits that goes as $g_1(t)g_2(t)\beta_m^1\beta_m^2/\tilde{\omega}_m$. A small frequency mismatch ($\delta_i-\tilde{\delta}_i$) in a coherent and close-to-resonant transfer of excitations leads to a residual population. Thus, the efficiency of state transfer can be estimated as $|z|^2\approx 1-\frac{(\delta_i-\tilde{\delta}_i)^2}{4g^2}$, for $|(\delta_i-\tilde{\delta}_i)/g|\ll 1$, where $g$ is the coupling between emitter and receptor. Approximating $\sum_m |\beta_m^i|^2/\tilde{\omega}_m\approx 1/\tilde{\omega}_0$ and $g_1(t)\approx \tilde{\kappa}$, we find $\frac{(\delta_i-\tilde{\delta}_i)^2}{4g^2}\approx 10^{-9}$ for $\kappa=2\pi\times 1$ MHz and $\eta=1$, which is consistent with the values found numerically (cf. Fig.~\ref{fig2}). It is worth stressing that this estimation is length independent, and predicts a  $1-|z|^2\sim \tilde{\kappa}^2=\kappa^2/\eta^2$ scaling (rather than $\kappa^4/\eta^4$ as in the distortion due to propagation in the short photon regime), which agrees with the results in the long-photon limit in Fig.~\ref{fig2}.

\section{Direct SWAP and improved STIRAP protocols}\label{app:swap}

\begin{figure}
    \includegraphics[width=\columnwidth]{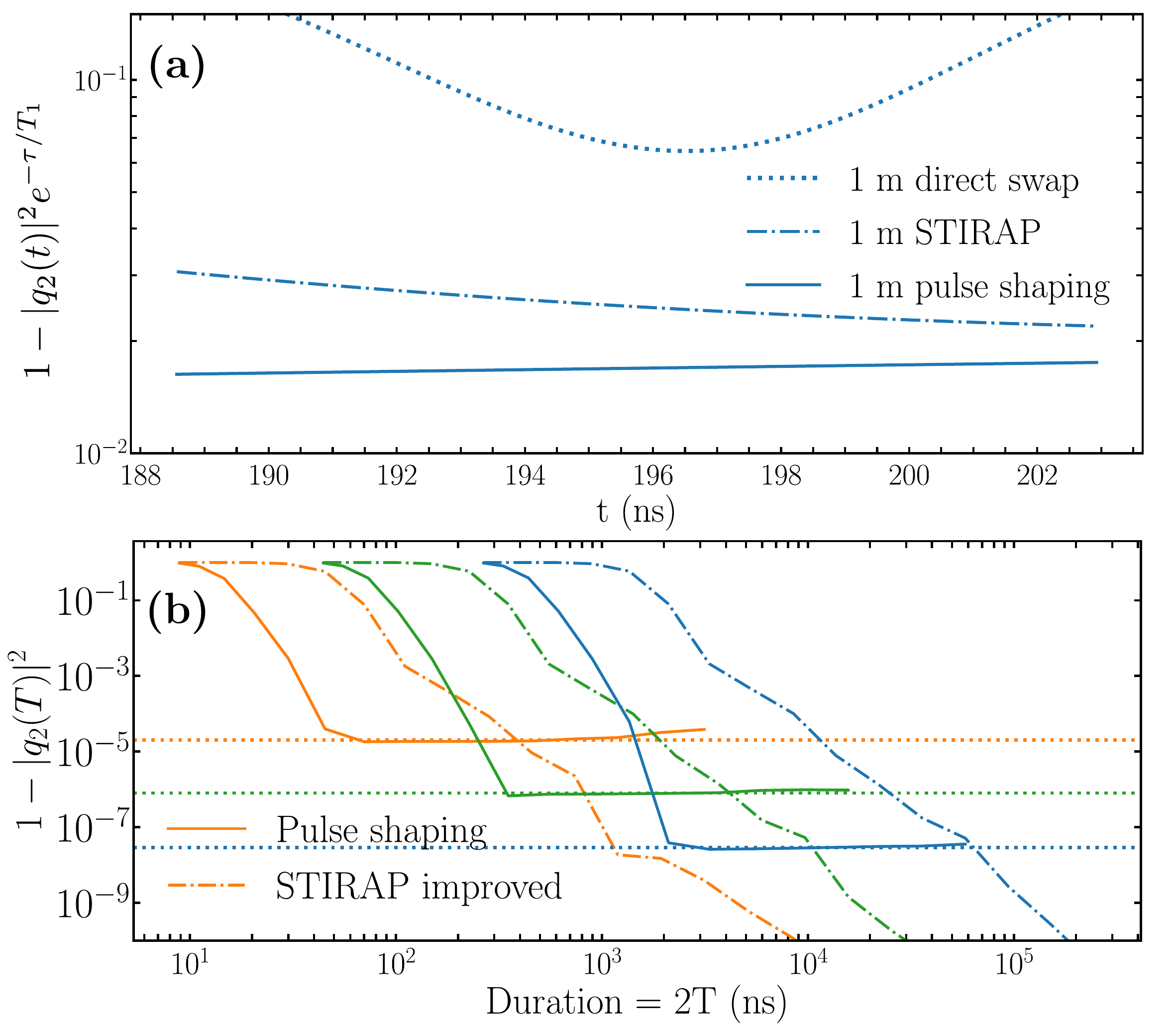}
\caption{(a) Robustness of each of the three protocols studied in this manuscript against slight variations of the protocol time for $\kappa = 2\pi\times 90 \ \text{MHz} \approx \Delta\omega_c$ and a $L=1$ m waveguide. The lines plotted correspond to a single simulation of those shown in Fig. ~\ref{fig3}(b) around the $200$ ns point. It can be seen that whereas the pulse shaping and the STIRAP (solid and dotted-dashed lines respectively) are robust, the direct SWAP (dotted line) undergoes large variations even for just a few nanoseconds. (b) Comparison of the quantum state transfer efficiency $1-|q_2(T)|^2$ as a function of the protocol duration for an improved STIRAP-like protocol (cf. App.~\ref{app:swap}) and pulse shaping following Eq.~\eqref{eq:gt}, for $L=1$, $5$, and $30$ m, as in Fig.~\ref{fig3}(a). The horizontal dotted lines correspond to the efficiency limit posed by the wavepacket distortion (cf. Eq.~\eqref{eq:stfidelity}), to which the pulse shaping saturates.}
\label{fig8}
\end{figure}

As reported in~\cite{Serafini2005}, a SWAP gate can be performed with an always-on and constant couplings $g_{1}$ and $g_2$. For simplicity, we consider $g_{1}=g_2=g$. When $g\ll G$, being $G$ the coupling between resonator and waveguide, the interaction between Q1 and Q2 is effectively mediated by a single mode, and hence the a quantum state transfer can be approximately achieved in a time $g t\approx \pi$~\cite{Serafini2005}. For a fixed coupling $g$, one must find the optimal time such that $|q_2(t)|^2$ is maximal. The results are plotted in Fig.~\ref{fig3}(a). In this manner, the fidelity of a direct-SWAP is sensitive to deviations from this optimal time, and thus less robust against this imperfection when compared to STIRAP-like or pulse-shaping protocols. We illustrate this fact for a particular example in Fig.~\ref{fig8}(a), which shows the dependence of $|q_2(t)|^2e^{-\tau/T_1}$, with $T_1=100\ \mu$s, in the vicinity of the optimal time, and compared it with the robust results for pulse shaping and STIRAP.

In Sec.~\ref{ss:stirap_swap} we have compared the performance of a quantum state transfer using pulse shaping with direct-SWAP and STIRAP-like protocols. As commented, the performance of the STIRAP can be improved when considering $g_1(t)=g_0\sin^2((t+T)\pi/(4T))$ and $g_2(t)=g_0 \cos^2((t+T)\pi/(4T))$ so that $dg_{1,2}(\pm T)/dt=0$. This is shown in Fig.~\ref{fig8}(b) for $g_0=\kappa/2$. Note that pulse shaping achieves better fidelities in a much shorter time.

\section{Fidelity limitations}\label{app:fid}
The fidelity of the quantum operations we consider are limited by the distortion. Here we provide a brief derivation of the expressions used in the main text.

First, the quantum state transfer operation can be written as the following unitary,
\begin{align}
    \mathcal{T}_{\rm id}=\begin{bmatrix} 1 & 0 &0 &0\\ 0 & 0 &1 &0\\ 0 & 1 &0 &0\\ 0 & 0 &0 &1\end{bmatrix}
\end{align}
in the basis $\{ \ket{00},\ket{01},\ket{10},\ket{11}\}$. In our case, we implement an imperfect $\mathcal{T}$. Since $\ket{00}$ remains unaltered under $H$ and $\ket{11}$ is not considered in our implementation, we are left with
\begin{align}
    F(U_\text{st}\ket{\psi},U\ket{\psi})=\left|1-|\beta|^2+|q_2(T)||\beta|^2 \right|^2
\end{align}
where $\ket{\psi}=\alpha \ket{00}+\beta\ket{01}$ with $|\alpha|^2+|\beta|^2=1$. In the previous expression, $q_2(T)$ is the amplitude of the state $\ket{1}$ of Q2 after the protocol. Thus, the quantum state transfer fidelity is obtained as $F_\text{st}=\min_{\beta}F(U_\text{st}\ket{\psi},U\ket{\psi})$~\cite{Nielsen}, which leads to $F_\text{st}=|q_2(T)|^2$.

Second, up to local single-qubit rotations, the unitary of the controlled-phase gate that we implement read as
\begin{align}
    U=\begin{bmatrix} 1 & 0 & 0 &0 \\ 0 &1 &0 &0\\ 0 & 0 & |z_{0}| &0\\ 0 &  0 & 0 & |z_1|e ^{-i \varphi} \end{bmatrix},
\end{align}
where $|z_1|=|z_0|=1$ and $\varphi=\pi$ corresponds to the ideal gate $U_{\rm id}$. We can effectively focus on the subspace spanned by $\{\ket{10},\ket{11} \}$. Here $|z_{x}|$ denotes  the Q1 amplitude after the protocol, namely,  $|q_{1,x}(T)|$ with $x=0,1$ stressing that it depends on the Q2 state, $\ket{0}$ or $\ket{1}$, respectively.
Considering a generic initial state, the minimum fidelity takes place for  a state of the form $\ket{\psi}=p_1 \ket{10}+\sqrt{1-|p_1|^2}\ket{11}$ with $|p_1|\leq 1$,
\begin{align}\label{eq:Fp1}
F(U_{\rm id}\ket{\psi},U \ket{\psi})=\left||p_1|^2|z_0|+(1-|p_1|^2)|z_1|e^{-i(\varphi-\varphi_{\rm id})}\right|^2. \nonumber
\end{align}
Let us denote by $F(p_1)$ the previous expression. Thus, minimum gate fidelity~\cite{Nielsen} follows from     $F_{\rm g}=\min_{p_1} F(p_1)$. Since $|z_1|\neq |z_0|$, we define $r=|z_0|/|z_1|$ with $0\leq r\leq 1$. The minimum over $p_1$ can be easily found, which results in Eq.~\eqref{eq:fidelity}. Note that if $r\approx 1$, then  $F_g\approx |z_0|^2\sin^2(\varphi/2)$.

Finally, we comment that the theoretical prediction for the distortion of the controlled phase gate can be derived from $z_x=\langle\tilde{\xi}_x(2t_p) |\xi_x(2t_p) \rangle$,
\begin{align}
    z_x= \int d\omega |f(\omega)|^2 e^{i\phi''_x(\omega)/2 (\omega-\omega_c)^2}e^{-it_p D/(v_g^2) (\omega-\omega_c)^2},
\end{align}
which is again equivalent to Eq.~\eqref{eq:Flimit} but setting $h=t_p D /(v_g^2)-\phi''_x(\omega_c)/2$. Note that $\phi''_x=(-1)^{x+1} 4/\kappa_2^2$, and thus $\phi''_1(\omega_c)>0$ reduces $h$, i.e. the scattering is able to mitigate part of the effect due to a non-linear dispersion relation, while for $\phi''_0(\omega_c)<0$ so both contributions sum up. This is the underlying reason for the asymmetry $|z_1|\geq |z_0|$. In particular, we can find the leading order contributions that spoil an ideal process,
\begin{align}\label{eq:ze_lc}
    |z_x|^2&\approx 1-\frac{\kappa_1^4}{45\eta^4}\left(\frac{ t_p D}{v_g^2} +(-1)^x\frac{2}{\kappa_2^2}\right)^2.
\end{align}
The phase correction to the ideal sought $\pi$ difference between $|\xi_1(2t_p)\rangle$ and $|\xi_0(2t_p)\rangle$ is given by ${\rm arg}\{z_1\}-{\rm arg}\{z_0 \}$. From the derivation given in App.~\ref{app:dis}, it follows
\begin{align}\label{eq:varphi_lc}
    \varphi\approx \pi+\frac{2\kappa_1^2}{3\kappa_2^2\eta^2}.
\end{align}


%

\end{document}